\documentclass[aps,prd,showpacs,twocolumn,10pt,floatfix,nofootinbib,runinaddress]{revtex4-1}
\usepackage{amssymb,amsfonts,amsmath,bm,bbm}
\usepackage{graphicx}
\usepackage{dcolumn}
\usepackage{bm}
\usepackage[
colorlinks=true,
linkcolor=blue, 
breaklinks=true,
urlcolor=blue,
citecolor=blue]{hyperref}
\usepackage{bbold}
\usepackage{epstopdf}

\newcommand{\ITP}{\affiliation{
CAS Key Laboratory of Theoretical Physics,
            Institute of Theoretical Physics, \\Chinese Academy of Sciences,
            Beijing 100190, China}}

\newcommand{\LZU}{\affiliation{
School of Nuclear Science and Technology, 
      Lanzhou University, Lanzhou 730000, China}}
      
\newcommand{\NBZJU}{\affiliation{
School of Information Science and Engineering, \\
      Zhejiang University Ningbo Institute of Technology, Ningbo 315100, China}}

\begin{document}

\title{Thermodynamics and susceptibilities of isospin imbalanced QCD matter}

\author{Zhen-Yan Lu} \email{luzhenyan@itp.ac.cn}\ITP

\author{Cheng-Jun~Xia} \email{cjxia@nit.zju.edu.cn}\NBZJU

\author{Marco Ruggieri} \email{ruggieri@lzu.edu.cn}\LZU

\date{\today}

\begin{abstract}

We study the thermodynamics and the susceptibilities of quark matter in the framework of two-flavor
Nambu$-$Jona-Lasinio model at finite isospin chemical potential and temperature.
Isospin number density, normalized energy density and trace anomaly are
shown to be in good agreement with 
the available lattice data as well as with the results from chiral perturbation theory at zero temperature. We also study how susceptibilities 
depend on the isospin chemical potential and on temperature. 
We find a peak for the chiral, pion, and isospin susceptibilities 
at the critical isospin chemical potential, $\mu_I^c(T)$, 
at the boundary of the phase transition between the normal and pion superfluid phase. 
Moreover, temperature makes the transition from  normal to  pion condensed
phase smoother.  
We also note that the pion susceptibility always remains zero in the normal phase while it is finite in the superfluid phase.

\end{abstract}

\maketitle

\section{Introduction} \label{INTRO}

There is compelling evidence that Quantum Chromodynamics (QCD) is the correct theory describing the strong interaction between quarks and gluons.
QCD is a non-Abelian gauge theory, which has a nontrivial vacuum state
and provides a rich phase structure and exciting 
physical phenomena in various environments~\cite{16Andersen.Naylor.ea25001-25001RMP,15Kharzeev193-214ARNPS,15Miransky.Shovkovy1-209PR,Philipsen-2019rjq}. At low energy densities quarks are confined inside  hadrons, 
while they form a weakly coupled plasma in the very high energy limit. 
Therefore, the strongly interacting QCD matter is expected to undergo a hadron-quark phase transition at high temperature and/or chemical potential. 
Moreover, the chiral symmetry is spontaneously broken at low temperature/baryon density
due to the nonvanishing chiral condensate, while it is expected to restore at high temperature
and high density~\cite{06Stephanov24-24P,11Fukushima.Hatsuda14001-14001RPP}. Lattice calculations show that at small baryon chemical potential the transition is actually a smooth crossover~\cite{Borsanyi:2010cj,Bazavov:2011nk,Borsanyi:2010bp,Borsanyi:2013bia,Cheng:2009zi}. 
On the other hand, many studies~\cite{98Alford.Rajagopal.ea247-256PLB,98Halasz.Jackson.ea96007-96007PRD,08Costa.Ruivo.ea96001-96001PRD} 
suggest that at large chemical potential
the smooth crossover meets a first order transition
at a particular point in the $(T,\mu)$ plane named the critical endpoint
(CEP), where the crossover becomes a second order 
phase transition. 
The quest for the possible existence of CEP as well as its precise location are still on-going, and have attracted a lot of attention, see e.g.
\cite{18Ferreira.Costa.ea34006-34006PRD,11Qin.Chang.ea172301-172301PRL,14Costa.Ferreira.ea56013-56013PRD,14Ruggieri.Oliva.ea255-260PLB,15Costa.Ferreira.ea36012-36012PRD,15Lu.Du.ea495-495EPJC,11Ruggieri14011-14011PRD}.

In QCD, fluctuations of the conserved charges like the baryon and isospin numbers, exhibit critical behavior around a phase transition;
these fluctuations are encoded into 
the corresponding susceptibilities
as well as into higher order moments~\cite{00Jeon.Koch2076-2079PRL,00Asakawa.Heinz.ea2072-2075PRL,03Hatta.Ikeda14028-14028PRD,03Hatta.Stephanov102003-102003PRL}.  
For example, the chiral susceptibility, which represents the modification of the chiral
condensate to a small perturbation of the current quark mass, 
is often used to describe the chiral phase transition~\cite{13Wang.Sun.ea1350064-1350064MPLA,08Min.Yu.ea76008-76008PRD,03Chakraborty.Mustafa.ea114004-114004PRD,08Zhao.Chang.ea483-492EPJC,05Bernard.Burch.ea34504-34504PRD,06Aoki.Fodor.ea46-54PLB,07Cheng.others34506-34506PRD}. 
Besides, it has been suggested that 
the moments of the baryon number can
be used to identify the CEP in the QCD phase
diagram~\cite{10Harada.Sasaki.ea16009-16009PRD,15Ding.Mukherjee.ea74043-74043PRD,03Fujii94018-94018PRD,04Fujii.Ohtani14016-14016PRD,91Kunihiro395-402PLB,98Stephanov.Rajagopal.ea4816-4819PRL,01Gavai.Gupta74506-74506PRD,02Gavai.Gupta.ea54506-54506PRD,11Cui.Takeuchi.ea76004-76004PRD,11Jiang.Luo.ea66-66J,09He.Li.ea36001-36001PRD,07Ratti.Roessner.ea57-60PLB}.
 
In the two-flavor case the quark chemical potentials $\mu_f$ can be 
expressed in term of the baryon chemical potential $\mu_B=N_c(\mu_u+\mu_d)/2$, and the isospin chemical potential $\mu_I=\mu_u-\mu_d$\footnote{Note that another definition of the isospin chemical potential $\mu_I=(\mu_u-\mu_d)/2$ is also frequently used in the literature, resulting in the critical point as $\mu_I^c(T=0)=m_\pi/2$ (see, e.g., Refs.~\cite{18Brandt.Endrfmmodeboxolseoidi.ea54514-54514PRD,Adhikari-2016vuu}).} with $N_c=3$ the number of colors. 
Along the $\mu_B$ axis, the knowledge from first principle lattice simulation on the thermodynamics of strongly interacting matter is limited, 
due to the notorious sign problem~\cite{03Muroya.Nakamura.ea615-668PTP}.
Thus, effective field theories of QCD as well as phenomenological models are necessary, and they might provide powerful tools for us to have a better understanding of QCD physics in the nonperturbative regime. On the other hand, lattice simulations are feasible for $\mu_B=0$ and
$\mu_I\neq0$, therefore it is possibe to study QCD at finite $\mu_I$ by means of first principle calculations and compare these with the 
predictions of effective models. This is an useful exercise to 
test the effectiveness of QCD models.
For example, an analysis based on the leading order (LO)~\cite{01Son.Stephanov592-595PRL,01Splittorff.Son.ea16003-16003PRD} and the next-to-leading order (NLO)~\cite{Adhikari-2019mdk,Adhikari-2019mlf} chiral perturbation theory (CHPT) shows that a second order phase transition from the normal to pion superfluid phase is expected to happen at $\mu_I^c(T=0)=m_\pi$, where $m_\pi$ is the pion mass. This has been confirmed by lattice simulations~\cite{02Kogut.Sinclair34505-34505PRD,02Kogut.Sinclair14508-14508PRD,04Kogut.Sinclair94501-94501PRD,12Detmold.Orginos.ea54507-54507PRD} as well as by the Nambu$-$Jona-Lasinio (NJL) model analytically~\cite{05He.Jin.ea116001-116001PRD,05He.Zhuang93-101PLB,05Warringa.Boer.ea14015-14015PRD}. 

Many aspects of pion condensate have been extensively studied in the effective field theories~\cite{03Loewe.Villavicencio74034-74034PRD,05Loewe.Villavicencio94001-94001PRD,15Adhikari.Cohen.ea45202-45202PRC,15Mammarella.Mannarelli85025-85025PRD,17Carignano.Lepori.ea35-35EPJA,16Carignano.Mammarella.ea51503-51503PRD,19Adhikari211-217PLB,Adhikari-2019zaj}, lattice simulations~\cite{16Brandt.Endrodi39-39P,13Shi12026-12026JPCS,18Scior.Smekal.ea7042-7042EWC,18Brandt.Endrodi.ea7020-7020EWC} and 
effective models~\cite{06Mao.Petropoulos.ea2187-2198JPG,10Matsuzaki16005-16005PRD,13Ueda.Nakano.ea74006-74006PRD,14Stiele.Fraga.ea72-78PLB,14Nishihara.Harada76001-76001PRD,09Andersen.Kyllingstad15003-15003JPG,13Xia.He.ea56013-56013PRD,17Wu.Ping.ea124106-124106CPC,18Chao.Huang.ea-,09Xiong.Jin.ea125005-125005JPG,Abuki:2008wm,Avancini-2019ego} (for a review, see \cite{Mannarelli-2019hgn}).
We also mention that  
a new type of compact star made of a charged pion condensate has
been proposed recently~\cite{14Mao116006-116006PRD,18Brandt.Endrodi.ea94510-94510PRD,18Andersen.Kneschke-}. 
While we refer to the original articles for the detailed picture for both two-flavor
and three-flavor quark matter, we remind here
the well established picture for the two-flavor case with finite bare quark masses. 
At zero temperature  there is a second order phase transition from the normal phase, in which
the pion condensate is vanishing, to the phase with nonzero pion condensate: this phase transition happens when $|\mu_I|=m_\pi$; increasing $|\mu_I|$
there is eventually another second order phase transition back to the normal phase. At finite temperature the gap between the two
critical $|\mu_I|$ becomes smaller and eventually the window 
for the pion condensate
phase closes: at very high temperature there is room for the normal
phase only,
see for example Fig.~4 of \cite{17Wu.Ping.ea124106-124106CPC}.

In this paper, we study several susceptibilities 
of isospin inbalanced QCD matter 
at vanishing $\mu_B$ and zero and/or finite temperature, 
which to the best of
our knowledge has not yet been explicitly discussed in the literature.
As mentioned above, doing this study is useful to test how effective 
models predictions compare with first principle calculations.
The thermodynamic properties of pion superfluid phase at vanishing temperature has been studied at the lowest order of CHPT at zero temperature~\cite{16Carignano.Mammarella.ea51503-51503PRD}
and it has been found a good agreement with the lattice QCD simulations 
for $\mu_I\lesssim 2m_\pi$, but also that the LO approximation breaks down at higher $\mu_I$. The NJL model can be 
employed in a larger range of chemical potential and temperature
with respect to LO CHPT and it is the purpose of this work
to show how this model describes the thermodynamics
and the fluctuations for the transition to the pion superfluid phase. 
We will use the two-flavor NJL model in this article for simplicity,
leaving the three-flavor case to a future study.

The article is organized as follows. In Sec.~\ref{NJLmuImuB}, we give a brief introduction for the theoretical
framework of NJL model at finite $\mu_I$ and temperature.
In Sec.~\ref{RESULT},
after comparing the NJL model results for several thermodynamic quantities with LO CHPT calculation and/or lattice data at vanishing temperature, we give the numerical results for several susceptibilities of our interest at finite $\mu_I$ and temperature.
Finally, we draw our conclusions in Sec.~\ref{CONCLUSION}

\section{The NJL model}\label{NJLmuImuB}

The standard NJL model Lagrangian density for two flavors of quarks is given by~\cite{61Nambu.Jona-Lasinio345-358PR,61Nambu.Jona-Lasinio246-254PR,91Vogl.Weise195-272PPNP,92Klevansky649-708RMP,93Volkov35-58PPN,94Hatsuda.Kunihiro221-367PR} 
\begin{eqnarray}\label{NJLLan}
\mathcal{L}=\bar{q}(i\gamma^\mu\partial_\mu-m+\hat{\mu}\gamma_0)q+G[(\bar{q}q)^2+(\bar{q}i\gamma_5\tau q)^2],
\end{eqnarray}
where $q$ represent the quark fields, $\tau$ the Pauli matrices
in flavor space, $G$ the four-fermion interaction coupling, 
$m$ the degenerate quark mass of up and down quarks and $\hat{\mu}=\mathrm{diag}\{\mu_u,\mu_d\}$ denotes the diagonal matrix for 
quark chemical potentials.

In the grand canonical ensemble
the thermodynamics of strongly interacting matter can be considered as a function of the isospin and baryon chemical potentials and  temperature. In this article we limit ourselves
to the case $\mu_B=0$ which leads to $\mu_u=\mu_I/2$, $\mu_d = -\mu_I/2$.
For simplicity,
we perform the calculations in the mean field approximation, i.e.,
\begin{eqnarray}
&&(\bar{q} q)^2 \approx  2(\bar{q} q)\langle \bar{q} q\rangle-\langle \bar{q} q\rangle^2,\\
&&(\bar{q}i\gamma_5\tau q)^2 \approx 
2(\bar{q}i\gamma_5\tau q)\langle \bar{q} i\gamma_5\tau q\rangle-\langle \bar{q} i\gamma_5\tau q\rangle^2.
\end{eqnarray}
We introduce the chiral condensates as
\begin{eqnarray}
\langle\bar{q}q\rangle=\sigma=\sigma_u+\sigma_d,
\end{eqnarray}
with $\langle\bar{u}u\rangle=\sigma_u$ and $\langle\bar{d}d\rangle=\sigma_d$.
At $\mu_I\neq 0$ a charged pion condensate is also expected: in order to account for this
we introduce
\begin{equation}
\pi^\pm =     \langle\bar{q}i\gamma_5\tau_\pm q \rangle \equiv  \Pi e^{\pm i \theta},
\end{equation}
with $\theta$ a real number and $\tau_\pm=\tau_1 \pm i\tau_2$; the effective potential
does not depend on $\theta$ so we choose $\theta=0$ in the above equation and we are left with
the pion condensates, $\Pi$, in the $\tau_1$-direction of isospin, namely  
\begin{eqnarray}
\Pi = \langle\bar{q}i\gamma_5\tau_1 q \rangle.
\end{eqnarray}
The Lagrangian density can thus be written as 
\begin{eqnarray}
\mathcal{L}&=&\bar{q}\big[i\gamma^\mu\partial_\mu-M_q+\hat{\mu}\gamma^0 \nonumber\\
&&+2G \Pi i\gamma^5\tau_1 \big]q
+G(\sigma^2+\Pi^2)
\end{eqnarray}
with the effective quark mass defined as $M_q=m-2G\sigma$. The
thermodynamic potential is
\begin{eqnarray}\label{POTENTIAL}
\Omega
&=&G(\sigma^2+\Pi^2)-\frac{2N_c}{\beta}\int\frac{d^3p}{(2\pi)^3} \nonumber\\
&&\times\Big\{\ln\left[(1+e^{-\beta E_p^+})(1+e^{\beta E_p^+})\right] \nonumber\\
&&+\ln\left[(1+e^{-\beta E_p^-})(1+e^{\beta E_p^-})\right]\Big\},
\end{eqnarray}
where $\beta=1/T$ and
\begin{eqnarray}
E_p&=&\sqrt{p^2+M^2},\\
E_p^{\pm}&=&\sqrt{\big(E_p\pm\mu_I/2\big)^2+4G^2\Pi^2}.
\end{eqnarray}
The ground state is given by the values of $\sigma$, $\Pi$ that minimize
$\Omega$: this is equivalent to require that $\sigma$ and $\Pi$ are
solution of the gap equations, 
\begin{eqnarray}\label{gapEQ}
\frac{\partial \Omega}{\partial \sigma}=\frac{\partial \Omega}{\partial \Pi}=0.  
\end{eqnarray}
  
Given $\Omega$ at the ground state it is straightforward to obtain the isospin number density, 
the pressure and the energy density as
\begin{eqnarray}
n_I&=&-\frac{\partial\Omega}{\partial\mu_I},\\
P&=&-(\Omega-\Omega_0), \label{NJL-P}\\
E&=&-P-T\frac{\partial \Omega}{\partial T}-\mu_I\frac{\partial \Omega}{\partial \mu_I},\label{NJL-E} 
\end{eqnarray}
where $\Omega_0$ is the thermodynamic potential  at $T=\mu_I=0$.

The two-flavor NJL model  has
three parameters: the current quark mass $m=0.005$ GeV, the four-fermion coupling strength $G=5.01$ GeV$^{-2}$, and the hard three-momentum cutoff $\Lambda=0.653$ GeV,
which are fixed by reproducing the empirical values of the pion mass $m_\pi=0.134$ GeV, the physical pion decay constant $f_\pi=0.093$ GeV, and the chiral condensate $\sigma_0=2(-0.25~\text{GeV})^3$ in vacuum~\cite{94Zhuang.Hufner.ea525-552NPA}.

Before going on it is useful to remind the results of LO CHPT
for the pressure and the energy density in the pion superfluid phase, namely
\begin{eqnarray}
P^{\text{LO}}&=&
\frac{f_\pi^2}{2\mu_I^2}(\mu_I^2-m_\pi^2)^2,\label{PressureCHPT}\\
E^{\text{LO}}&=&\frac{f_\pi^2}{2\mu_I^2}(\mu_I^4+2\mu_I^2m_\pi^2-3m_\pi^4), \label{EnergyCHPT}
\end{eqnarray}
where $f_\pi$ is the pion decay constant.
The result for isospin density will be also useful in the following section:
\begin{eqnarray}\label{nICHPT}
n_I=\frac{\partial P^{\text{LO}}}{\partial \mu_I}=\frac{f_\pi^2}{\mu_I^3}(\mu_I^4-m_\pi^4).
\end{eqnarray}

\section{Results and discussion}\label{RESULT}

\subsection{Isospin number density, energy density, and trace anomaly}

\begin{figure}
  \includegraphics[width=246pt]{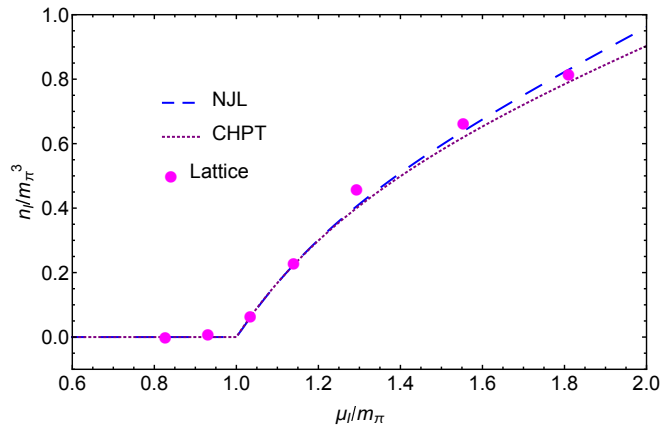}\\
  \caption{The normalized isospin number density as a function of $\mu_I/m_\pi$ at fixed $T=0$. The blue dashed line
  corresponds to the result obtained from the NJL model. The result from the CHPT (purple dotted line), i.e., Eq.~(\ref{nICHPT}), and recent lattice data~\cite{18Brandt.Endrodi.ea94510-94510PRD} (magenta circles) are also included for comparison.  
 }\label{nI-muI-T0}
\end{figure}

In Fig.~\ref{nI-muI-T0}, we plot the isospin number density 
normalized to $m_\pi^3$ at zero temperature as a function of $\mu_I/m_\pi$.
For comparison we also include the results 
from LO CHPT\footnote{Recently, the contribution from Lee-Huang-Yang term is also included~\cite{19Lepori.Mannarelli96011-96011PRD}.} (purple dotted line) as well as recent lattice data (magenta circles)~\cite{18Brandt.Endrodi.ea94510-94510PRD}.
The results obtained from these methods agree quantitatively
with each other in the considered range of $\mu_I$. 
When $\mu_I<m_\pi$ the isospin number density
is zero, which corresponds to the normal phase of the system. 
In the pion superfluid phase with $\mu_I> m_\pi$, the isospin number
density becomes nonzero, which increases monotonically with $\mu_I$.

\begin{figure}
  \includegraphics[width=246pt]{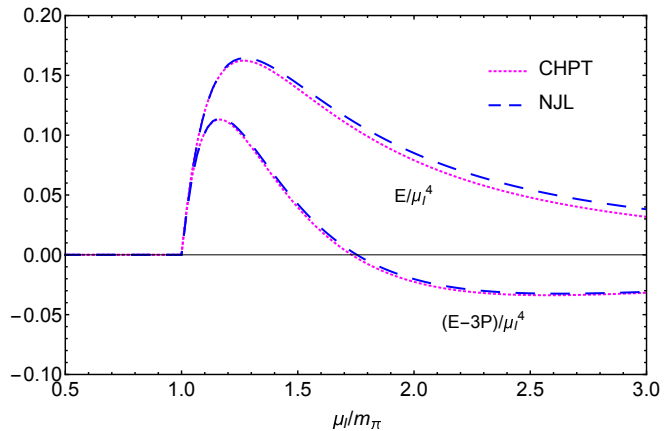}\\
  \caption{The normalized energy density $E/\mu_I^4$ and trace anomaly $(E-3P)/\mu_I^4$ as functions of $\mu_I/m_\pi$ at zero temperature.
  The result from CHPT at zero temperature is also included for comparison.}\label{EmuI4-muI}
\end{figure}

In Fig.~\ref{EmuI4-muI}, we plot the energy density  normalized to $\mu_I^4$ as a function of $\mu_I/m_\pi$ 
at $T=0$. The results from the LO CHPT (purple dotted line) in Eq.~(\ref{EnergyCHPT}) 
are also shown for comparison.
The result from the NJL model agrees with 
the one from CHPT at zero temperature though the energy density from the former is slightly larger than that from the latter at high $\mu_I$.
The quantity $E/\mu_I^4$
becomes nonzero and positive when $\mu_I$ exceeds the critical value $\mu_I^c(T=0)=m_\pi$. 
In particular, it develops a peak at $\mu_I^{\text{peak}}\simeq 1.274m_\pi$. Recently, a calculation of this peak based on the LO CHPT has been
done in Ref.~\cite{16Carignano.Mammarella.ea51503-51503PRD};
moreover, this quantity has also been computed on the lattice:
\begin{eqnarray}
\mu_I^{\text{peak}}\simeq
\begin{cases}
1.274m_\pi,  &\text{NJL},\cr
1.276m_\pi,  &\text{CHPT}~\text{\cite{16Carignano.Mammarella.ea51503-51503PRD}},\cr
\{1.20,1.25,1.275\}m_\pi,  &\text{Lattice data}~\text{\cite{12Detmold.Orginos.ea54507-54507PRD}}.
\end{cases}
\end{eqnarray}
In the above equation the three lattice results quoted have been
obtained with three different lattice volumes,
namely $L^3$=$\{16^3$, $20^3$, $24^3\}$ respectively.
The extrapolation of the lattice results to the continuum limit
gives $\mu_I^{\text{peak}}=1.30(7)m_\pi$.
We notice the nice agreement of the NJL model with LO CHPT and lattice data.

In Fig.~\ref{EmuI4-muI} we also show the normalized trace anomaly,  $(E-3P)/\mu_I^4$, as a function of $\mu_I/m_\pi$ at $T=0$. 
Again we notice the agreement between the NJL model and the LO CHPT 
results in the entire isospin chemical potential range considered. 
In more detail, we observe that the blue dashed line (lower one), which represents the result calculated
in the NJL model at zero temperature, keeps zero at small
$\mu_I$, and becomes nonzero and positive when $\mu_I$ slightly larger than $m_\pi$.
Similar to the behavior of $E/\mu_I^4$, the normalized trace anomaly also develops a peak at an intermediate $\mu_I$ although 
the peak of the former
appears at a slightly smaller value of $\mu_I$
with respect to the latter. 
As $\mu_I$ increases inside the domain of the pion
condensed phase, the normalized trace anomaly decreases and 
becomes negative.
In fact, the point $\bar{\mu}_I$
fulfilling the conformal relation $E-3P=0$ can be evaluated in the NJL model at zero temperature as
\begin{eqnarray}
\bar{\mu}_I \simeq 1.754m_\pi,
\end{eqnarray}
which is in good agreement with the LO CHPT result given by $\bar{\mu}_I=\sqrt{3}m_\pi$~\cite{16Carignano.Mammarella.ea51503-51503PRD}.
This point separates the $E>3P$ and $E<3P$ regions. 
In fact, this value $\bar{\mu}_I \simeq 1.754m_\pi$ is very close to the value of $\mu_I$ corresponding to the BEC-BCS crossover, which is estimated to be $1.702m_\pi$ at zero temperature according to the analyses in Refs.~\cite{He-2006tn,07Sun.He.ea96004-96004PRD,07He.Zhuang96003-96003PRD,Abuki-2010jq}. However, although these two values almost coincide with each other, whether $\bar{\mu}_I$ can be identified with the isospin chemical potential corresponding to the BEC-BCS crossover or not need to be investigated in more detail in the future.


\subsection{Chiral, pion, and isospin susceptibilities}

\begin{figure}
  \includegraphics[width=246pt]{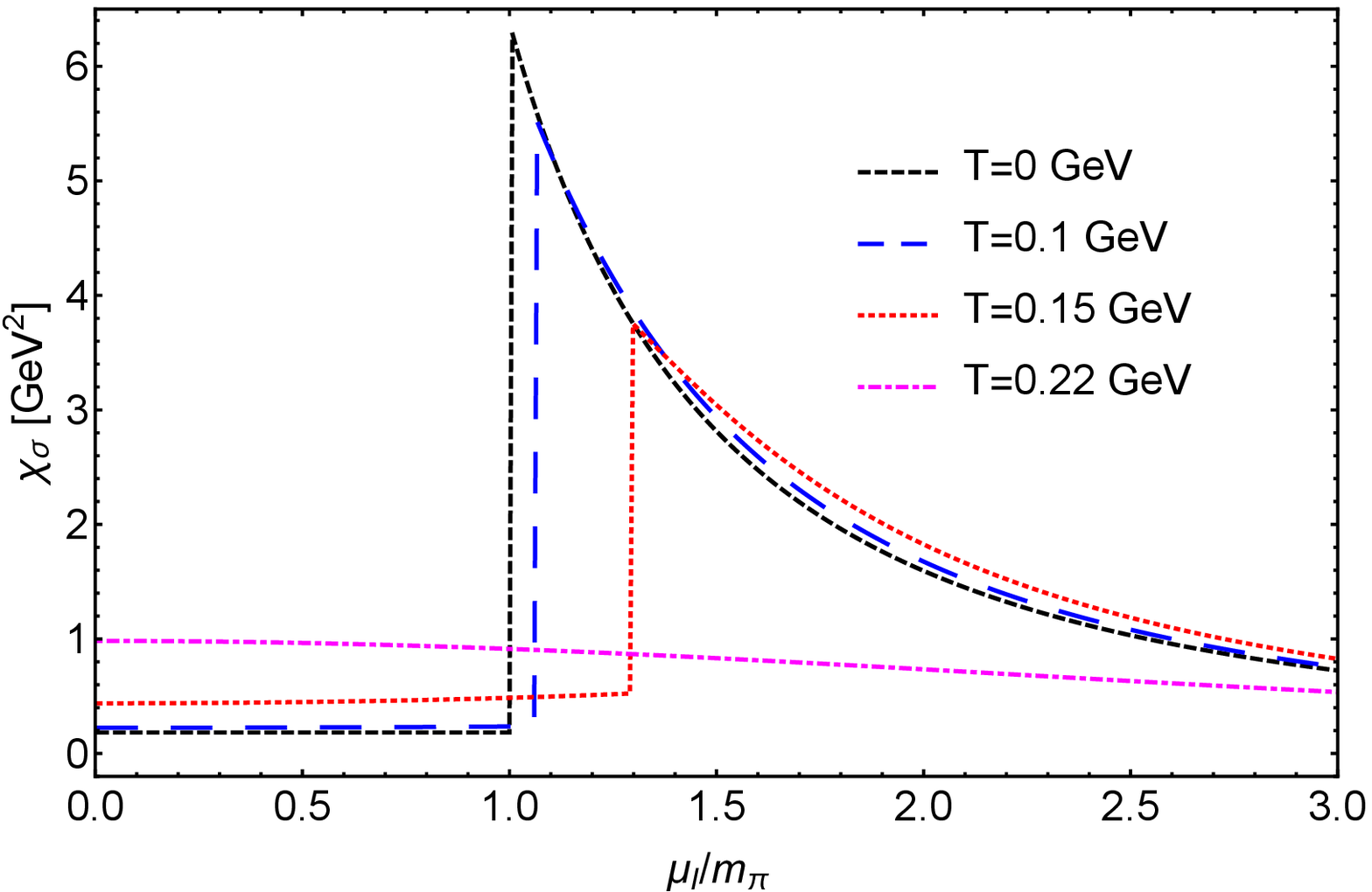}\\
    \includegraphics[width=246pt]{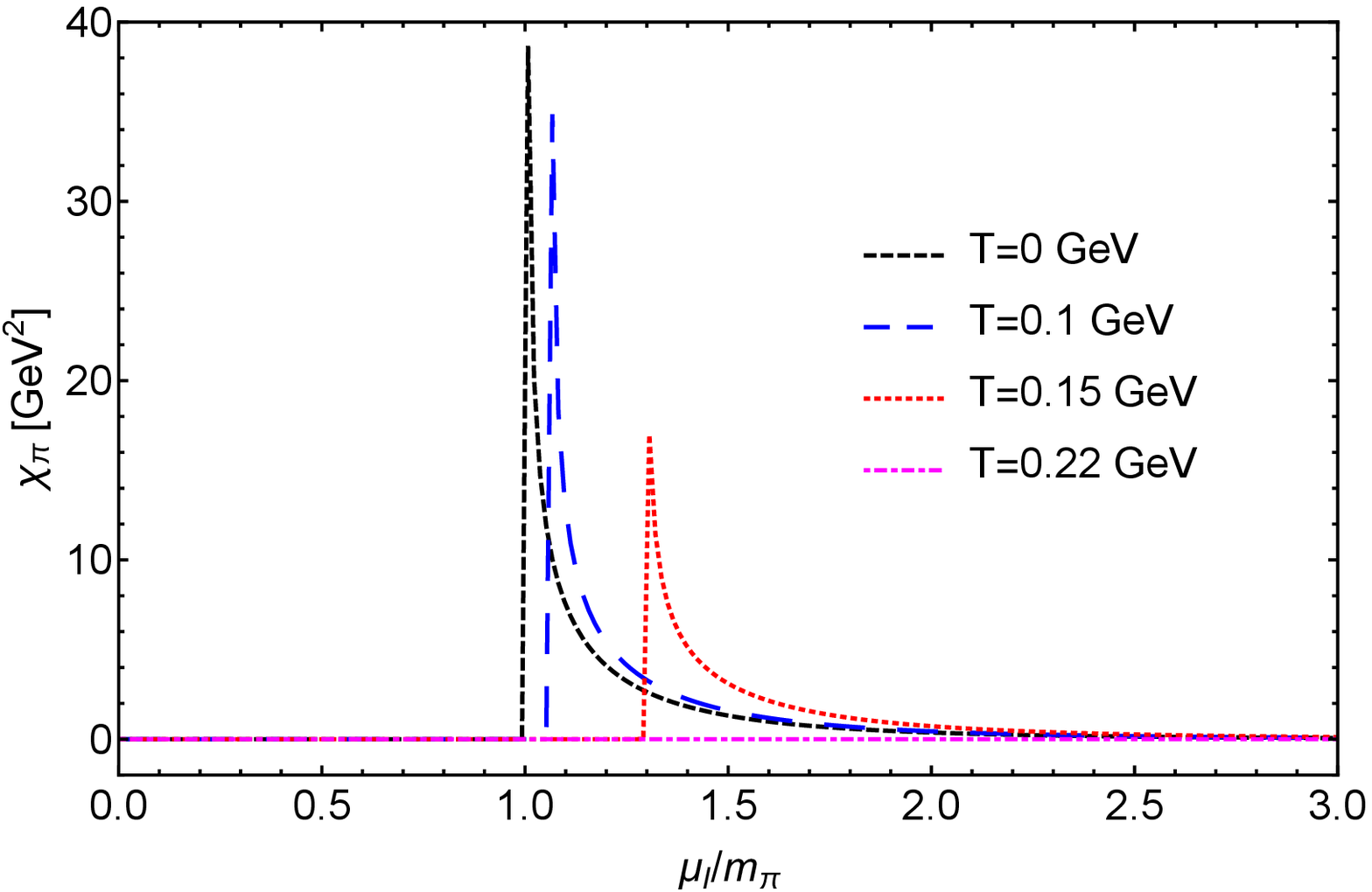}\\
      \includegraphics[width=246pt]{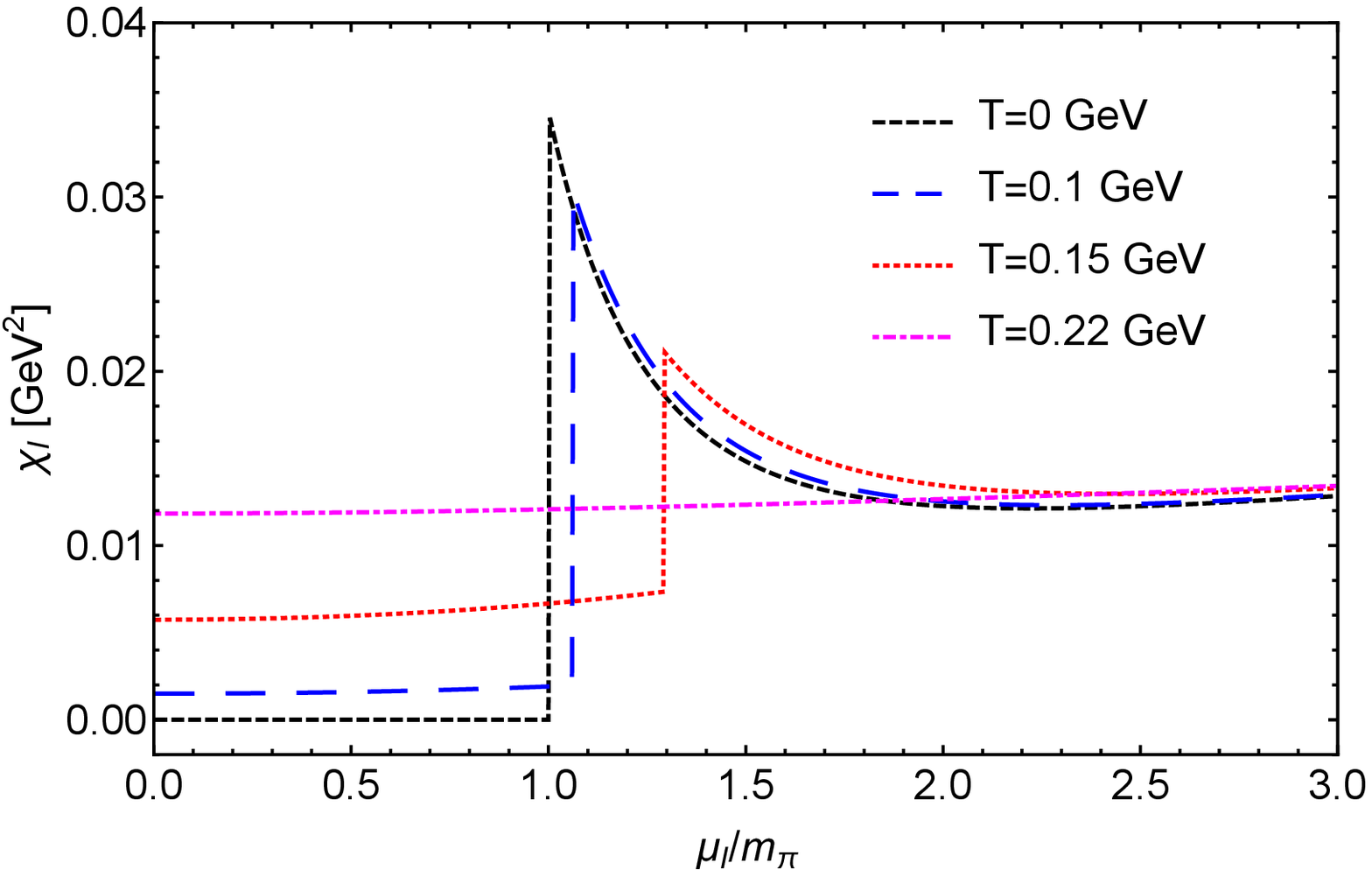}\\
  \caption{The variation behaviors of chiral susceptibility $\chi_\sigma$, pion susceptibility $\chi_\pi$, and isospin number susceptibility $\chi_I$ with respect to $\mu_I/m_\pi$ for several values of the temperature. }\label{chiSIGMA-muI}
\end{figure}

The transition to the pion condensate  phase can be well identified 
by means of the susceptibilities. 
In principle, they  can also be measured on the lattice since there is no sign problem at finite $\mu_I$
due to the real and positive fermionic determinant.
Chiral condensate is the order parameter of spontaneous chiral symmetry breaking, while pion condensate indicates the spontaneous
isospin symmetry breaking. The chiral susceptibility, $\chi_\sigma$, which corresponds to the zero-momentum projection of the scalar
propagator and encodes all fluctuations of the order parameter, is defined as~\cite{94Karsch.Laermann6954-6962PRD,05Bernard.Burch.ea34504-34504PRD,03Chakraborty.Mustafa.ea114004-114004PRD,06Aoki.Fodor.ea46-54PLB}
\begin{eqnarray}
\chi_\sigma=-\frac{\partial \sigma}{\partial m}. 
\end{eqnarray}
We can also define the pion susceptibility, which is obtained as the first derivative of pion condensate with respect to the current quark mass, i.e., \begin{eqnarray}
\chi_\pi=\frac{\partial\Pi}{\partial m}.
\end{eqnarray}
We also define the isospin number susceptibility as
\begin{eqnarray}
\chi_I=\frac{d n_I}{d\mu_I}.
\end{eqnarray} 
Note that we write the derivative operator above as a total derivative instead of a partial one because one might take into account the fact that
the condensates may have a dependence on $\mu_I$.

The chiral, pion, and isospin susceptibilities as functions of $\mu_I/m_\pi$ for four values of the temperature are
illustrated in Fig.~\ref{chiSIGMA-muI}. The temperatures are chosen as $T=0$ (black short-dashed line), $T=0.1$ GeV (blue dashed line), $T=0.15$ GeV (red dotted line), and $T=0.22$ GeV (magenta dot-dashed line). 
We notice that the zero temperature susceptibilities
exhibit a discontinuity at $\mu_I=\mu_I^c(T)$, signaling the boundary of a second order phase transition.
At nonzero temperatures, e.g., 
$T=0.1$ GeV and 0.15 GeV, the situation is similar except that the discontinuity shifts to larger $\mu_I$.
For $T=0.22$ GeV instead  the susceptibilities are continuous in the whole range of $\mu_I$, which can be easily understood with the
absence of a transition to the pion condensed phase
in agreement with \cite{05He.Jin.ea116001-116001PRD}.
We notice that the chiral susceptibility stays finite in the
pion condensed phase, while the pion susceptibility vanishes
in the normal phase: this is obviously related to the fact that
the bare quark mass is finite, so chiral symmetry is always broken
explicitly and this leads to a finite chiral condensate.
In the chiral limit this would not happen and the chiral susceptibility
would vanish in the pion condensed phase.

\begin{figure}
  \includegraphics[width=246pt]{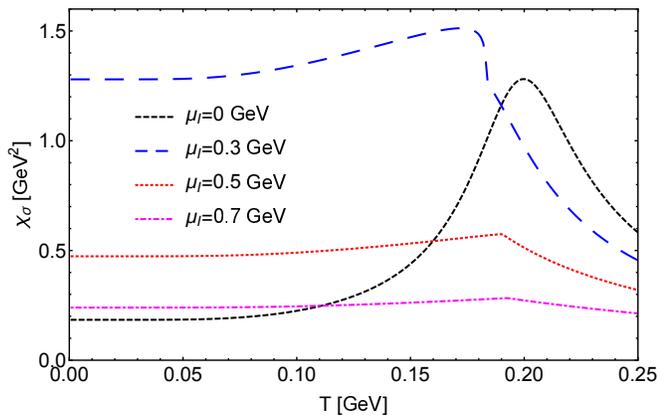}\\
  \caption{The variation behaviors of the chiral susceptibility with respect to temperature at several values of $\mu_I$. }\label{chiSIGMA-T}
\end{figure}

In Fig.~\ref{chiSIGMA-T}, we show the thermal behavior of the chiral susceptibility at several
values of $\mu_I$. The chiral susceptibility at $\mu_I=0$, 
which is denoted by the black short-dashed line, firstly grows up
smoothly at low temperature, then it develops a smooth peak at 
$T=0.2$ GeV in correspondence of the chiral crossover. 
At higher temperatures, the chiral
susceptibility decreases with temperature. 
We also include several cases with $\mu_I>\mu_I^c(T)$ for comparison. For $\mu_I=0.3$ GeV,
which is represented by the blue dashed line, the chiral susceptibility has a much larger value at $T=0$ compared to the other cases.
This line first slowly increases with temperature and reaches a smooth peak at $\mu_I/m_\pi=0.175$, signaling a crossover.      
After the peak the line drops and show a  kink at $\mu_I/m_\pi=0.183$, in correspondence of the 
transition to the normal phase.  
Similarly, the lines for $\mu_I=0.5$ GeV and 0.7 GeV, which are represented by red dotted and orange dot-dashed lines respectively,
also show a modest kink peak at the critical temperature.

\begin{figure}
  \includegraphics[width=246pt]{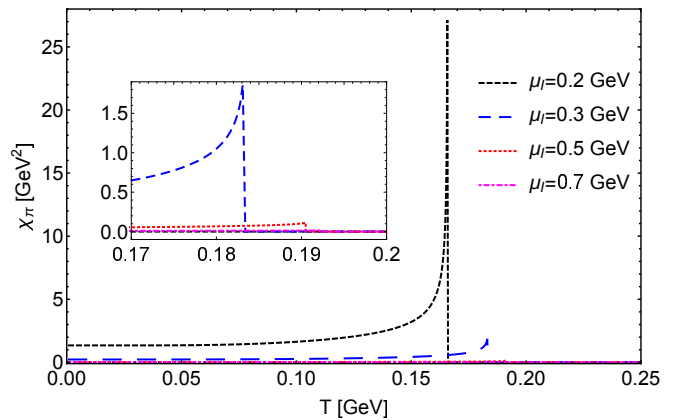}\\
  \caption{The variation behaviors of the pseudoscalar susceptibility against temperature for several values of $\mu_I$. }\label{chiPIC-T}
\end{figure}

In Fig.~\ref{chiPIC-T}, we plot $\chi_\pi$ versus temperature
for several values of $\mu_I$, i.e., $\mu_I=0.2,~0.3,~0.5,~0.7$ GeV.
We do not illustrate the result for $\chi_\pi$ with $\mu_I$ smaller than the critical $\mu_I^c(T)$
because in the normal phase $\chi_\pi=0$. 
For $\mu_I=0.2$ GeV, which is labeled by the black short-dashed line, $\chi_\pi$ increases very smoothly at low temperature, and then
grows up rapidly when $T$ is close to the critical temperature; 
clearly it vanishes above this temperature because the pion
condensate is zero there.

\begin{figure}
  \includegraphics[width=246pt]{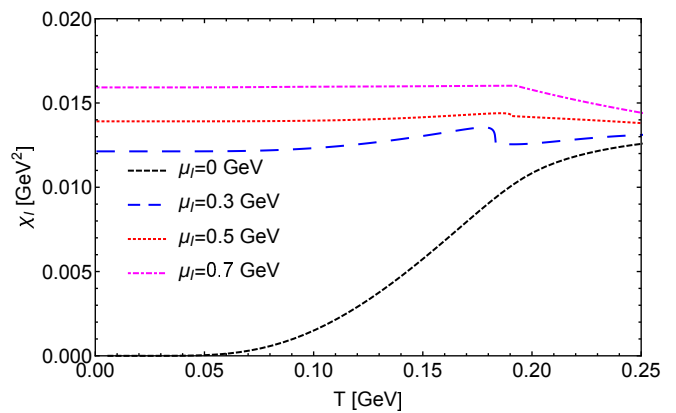}\\
  \caption{The isospin number susceptibility $\chi_I$ as a function of the temperature for several values of $\mu_I$. }\label{chiImuImPi}
\end{figure}

In Fig.~\ref{chiImuImPi} We plot $\chi_I$ versus the temperature for several values of $\mu_I$ . For the black short-dashed line, which represents the result obtained at $\mu_I=0$, the isospin number susceptibility increases smoothly and monotonously with temperature. At higher $\mu_I$, e.g., $\mu_I=0.3$ GeV (blue dashed line), $\mu_I=0.5$ GeV (red dotted line), and $\mu_I=0.7$ GeV (magenta dot-dashed line), $\chi_I$ keep almost unchanged in the entire considered temperature range expect for the cusp at the phase transition point.

\section{Conclusions}  \label{CONCLUSION}

In this work, we have studied the isospin chemical potential and temperature dependence of several thermodynamic quantities, in particular of several susceptibilities, of isospin-imbalanced QCD matter at vanishing baryon chemical potential 
within the two-flavor NJL model. We have compared the  isospin number density at zero temperature with recent lattice data as well as with LO CHPT, finding  a good agreement with the latter results, see Fig.~\ref{nI-muI-T0}. In addition, the result for the normalized energy density $E/\mu_I^4$ as a function of $\mu_I/m_\pi$ is also in fair agreement with that from LO CHPT at zero temperature. 
In particular, the location of the peak of $E/\mu_I^4$ as well as the trace anomaly computed within the NJL model agree with LO CHPT and/or lattice simulations: we have found  
$\mu_I^{\text{peak}}\simeq 1.274m_\pi$ for the peak of the energy density
and $\bar{\mu}_I \simeq 1.754m_\pi$ for the condition $E-3P=0$.

We have then considered the chiral, pion and isospin susceptibilities
at finite temperature. These are interesting quantities because they
allow to define the critical lines in the $(T,\mu_I)$ plane and
can also be computed on the lattice, so they can be used as a test
of the effective models of QCD.
The qualitative behavior of these susceptibilities is in agreement
with naive expectations: in particular, the chiral susceptibility
is always finite because of the nonvanishing bare quark mass
that induces a finite chiral condensate even at large $T/\mu_I$.
On the other hand, the pion susceptibility vanishes in the normal phase
and is nonzero in the pion condensed phase.
The transition from the normal phase to the pion condensed phase is
of the second order with a divergent pion susceptibility.
The isospin number susceptibility also keep finite and almost unchanged in the normal phase except that it is zero for the zero temperature case,
and shows a discontinuity at the phase transition. 

In this study we have ignored the role of 
baryon and 
strangeness chemical potentials~\cite{16Carignano.Mammarella.ea51503-51503PRD}.  These external sources are known to play an important role and we leave the study of the more complete problem to future works.

\section*{Acknowledgement}
The authors acknowledge Jingyi Chao and John Petrucci for inspiration
and comments on the first version of this article.
ZYL acknowledges the warm hospitality of 
the Department of Nuclear Science and Technology of Lanzhou University where part of this work
was done.
This  work  is  supported  in  part  by  the  China  Postdoctoral Science Foundation under Grant No. 2017M620920, by DFG and
NSFC through funds provided to the Sino-German
CRC 110 ``Symmetries and the Emergence of Structure
in QCD'' (NSFC Grant No. 11621131001 and DFG
Grant No.~TRR110), by
the NSFC under Grant Nos. 11747601, 11835015, and 11705163,
by the Chinese Academy of Sciences (CAS) under Grant No. QYZDB-SSW-SYS013 and No. XDPB09, and
by the CAS Center for Excellence in Particle Physics (CCEPP).
The work of M. R. is supported by the National Science Foundation of China (Grants No.~11805087 and No. 11875153)
and by the Fundamental Research Funds for the Central Universities (Grants No.~862946).

\bibliographystyle{aapmrev4-2}  
\bibliography{MyRefLu}

\begin{thebibliography}{99}%
\makeatletter
\providecommand \@ifxundefined [1]{%
 \@ifx{#1\undefined}
}%
\providecommand \@ifnum [1]{%
 \ifnum #1\expandafter \@firstoftwo
 \else \expandafter \@secondoftwo
 \fi
}%
\providecommand \@ifx [1]{%
 \ifx #1\expandafter \@firstoftwo
 \else \expandafter \@secondoftwo
 \fi
}%
\providecommand \natexlab [1]{#1}%
\providecommand \enquote  [1]{``#1''}%
\providecommand \bibnamefont  [1]{#1}%
\providecommand \bibfnamefont [1]{#1}%
\providecommand \citenamefont [1]{#1}%
\providecommand \href@noop [0]{\@secondoftwo}%
\providecommand \href [0]{\begingroup \@sanitize@url \@href}%
\providecommand \@href[1]{\@@startlink{#1}\@@href}%
\providecommand \@@href[1]{\endgroup#1\@@endlink}%
\providecommand \@sanitize@url [0]{\catcode `\\12\catcode `\$12\catcode
  `\&12\catcode `\#12\catcode `\^12\catcode `\_12\catcode `\%12\relax}%
\providecommand \@@startlink[1]{}%
\providecommand \@@endlink[0]{}%
\providecommand \url  [0]{\begingroup\@sanitize@url \@url }%
\providecommand \@url [1]{\endgroup\@href {#1}{\urlprefix }}%
\providecommand \urlprefix  [0]{URL }%
\providecommand \Eprint [0]{\href }%
\providecommand \doibase [0]{https://doi.org/}%
\providecommand \selectlanguage [0]{\@gobble}%
\providecommand \bibinfo  [0]{\@secondoftwo}%
\providecommand \bibfield  [0]{\@secondoftwo}%
\providecommand \translation [1]{[#1]}%
\providecommand \BibitemOpen [0]{}%
\providecommand \bibitemStop [0]{}%
\providecommand \bibitemNoStop [0]{.\EOS\space}%
\providecommand \EOS [0]{\spacefactor3000\relax}%
\providecommand \BibitemShut  [1]{\csname bibitem#1\endcsname}%
\let\auto@bib@innerbib\@empty
\bibitem [{\citenamefont {Andersen}, \citenamefont {Naylor},\ and\
  \citenamefont {Tranberg}(2016)}]{16Andersen.Naylor.ea25001-25001RMP}%
  \BibitemOpen
  \bibfield  {author} {\bibinfo {author} {\bibfnamefont {J.~O.}\ \bibnamefont
  {Andersen}}, \bibinfo {author} {\bibfnamefont {W.~R.}\ \bibnamefont
  {Naylor}},\ and\ \bibinfo {author} {\bibfnamefont {A.}~\bibnamefont
  {Tranberg}},\ }\href {https://doi.org/10.1103/RevModPhys.88.025001}
  {\bibfield  {journal} {\bibinfo  {journal} {Rev. Mod. Phys.}\ }\textbf
  {\bibinfo {volume} {88}},\ \bibinfo {pages} {025001} (\bibinfo {year}
  {2016})},\ \Eprint {https://arxiv.org/abs/1411.7176} {arXiv:1411.7176
  [hep-ph]} \BibitemShut {NoStop}%
\bibitem [{\citenamefont {Kharzeev}(2015)}]{15Kharzeev193-214ARNPS}%
  \BibitemOpen
  \bibfield  {author} {\bibinfo {author} {\bibfnamefont {D.~E.}\ \bibnamefont
  {Kharzeev}},\ }\href {https://doi.org/10.1146/annurev-nucl-102313-025420}
  {\bibfield  {journal} {\bibinfo  {journal} {Ann. Rev. Nucl. Part. Sci.}\
  }\textbf {\bibinfo {volume} {65}},\ \bibinfo {pages} {193} (\bibinfo {year}
  {2015})},\ \Eprint {https://arxiv.org/abs/1501.01336} {arXiv:1501.01336
  [hep-ph]} \BibitemShut {NoStop}%
\bibitem [{\citenamefont {Miransky}\ and\ \citenamefont
  {Shovkovy}(2015)}]{15Miransky.Shovkovy1-209PR}%
  \BibitemOpen
  \bibfield  {author} {\bibinfo {author} {\bibfnamefont {V.~A.}\ \bibnamefont
  {Miransky}}\ and\ \bibinfo {author} {\bibfnamefont {I.~A.}\ \bibnamefont
  {Shovkovy}},\ }\href {https://doi.org/10.1016/j.physrep.2015.02.003}
  {\bibfield  {journal} {\bibinfo  {journal} {Phys. Rept.}\ }\textbf {\bibinfo
  {volume} {576}},\ \bibinfo {pages} {1} (\bibinfo {year} {2015})},\ \Eprint
  {https://arxiv.org/abs/1503.00732} {arXiv:1503.00732 [hep-ph]} \BibitemShut
  {NoStop}%
\bibitem [{\citenamefont {Philipsen}(2019)}]{Philipsen-2019rjq}%
  \BibitemOpen
  \bibfield  {author} {\bibinfo {author} {\bibfnamefont {O.}~\bibnamefont
  {Philipsen}}\ }(\bibinfo {year} {2019})\ \Eprint
  {https://arxiv.org/abs/1912.04827} {arXiv:1912.04827 [hep-lat]} \BibitemShut
  {NoStop}%
\bibitem [{\citenamefont {Stephanov}(2006)}]{06Stephanov24-24P}%
  \BibitemOpen
  \bibfield  {author} {\bibinfo {author} {\bibfnamefont {M.~A.}\ \bibnamefont
  {Stephanov}},\ }\bibfield  {booktitle} {\emph {\bibinfo {booktitle}
  {{Proceedings, 24th International Symposium on Lattice Field Theory (Lattice
  2006): Tucson, USA, July 23-28, 2006}}},\ }\href
  {https://doi.org/10.22323/1.032.0024} {\bibfield  {journal} {\bibinfo
  {journal} {PoS}\ }\textbf {\bibinfo {volume} {LAT2006}},\ \bibinfo {pages}
  {024} (\bibinfo {year} {2006})},\ \Eprint
  {https://arxiv.org/abs/hep-lat/0701002} {arXiv:hep-lat/0701002 [hep-lat]}
  \BibitemShut {NoStop}%
\bibitem [{\citenamefont {Fukushima}\ and\ \citenamefont
  {Hatsuda}(2011)}]{11Fukushima.Hatsuda14001-14001RPP}%
  \BibitemOpen
  \bibfield  {author} {\bibinfo {author} {\bibfnamefont {K.}~\bibnamefont
  {Fukushima}}\ and\ \bibinfo {author} {\bibfnamefont {T.}~\bibnamefont
  {Hatsuda}},\ }\href {https://doi.org/10.1088/0034-4885/74/1/014001}
  {\bibfield  {journal} {\bibinfo  {journal} {Rept. Prog. Phys.}\ }\textbf
  {\bibinfo {volume} {74}},\ \bibinfo {pages} {014001} (\bibinfo {year}
  {2011})},\ \Eprint {https://arxiv.org/abs/1005.4814} {arXiv:1005.4814
  [hep-ph]} \BibitemShut {NoStop}%
\bibitem [{\citenamefont {Borsanyi}\ \emph
  {et~al.}(2010{\natexlab{a}})\citenamefont {Borsanyi}, \citenamefont
  {Endrodi}, \citenamefont {Fodor}, \citenamefont {Jakovac}, \citenamefont
  {Katz}, \citenamefont {Krieg}, \citenamefont {Ratti},\ and\ \citenamefont
  {Szabo}}]{Borsanyi:2010cj}%
  \BibitemOpen
  \bibfield  {author} {\bibinfo {author} {\bibfnamefont {S.}~\bibnamefont
  {Borsanyi}}, \bibinfo {author} {\bibfnamefont {G.}~\bibnamefont {Endrodi}},
  \bibinfo {author} {\bibfnamefont {Z.}~\bibnamefont {Fodor}}, \bibinfo
  {author} {\bibfnamefont {A.}~\bibnamefont {Jakovac}}, \bibinfo {author}
  {\bibfnamefont {S.~D.}\ \bibnamefont {Katz}}, \bibinfo {author}
  {\bibfnamefont {S.}~\bibnamefont {Krieg}}, \bibinfo {author} {\bibfnamefont
  {C.}~\bibnamefont {Ratti}},\ and\ \bibinfo {author} {\bibfnamefont {K.~K.}\
  \bibnamefont {Szabo}},\ }\href {https://doi.org/10.1007/JHEP11(2010)077}
  {\bibfield  {journal} {\bibinfo  {journal} {JHEP}\ }\textbf {\bibinfo
  {volume} {11}},\ \bibinfo {pages} {077} (\bibinfo {year}
  {2010}{\natexlab{a}})},\ \Eprint {https://arxiv.org/abs/1007.2580}
  {arXiv:1007.2580 [hep-lat]} \BibitemShut {NoStop}%
\bibitem [{\citenamefont {Bazavov}\ \emph {et~al.}(2012)\citenamefont {Bazavov}
  \emph {et~al.}}]{Bazavov:2011nk}%
  \BibitemOpen
  \bibfield  {author} {\bibinfo {author} {\bibfnamefont {A.}~\bibnamefont
  {Bazavov}} \emph {et~al.},\ }\href
  {https://doi.org/10.1103/PhysRevD.85.054503} {\bibfield  {journal} {\bibinfo
  {journal} {Phys. Rev. D}\ }\textbf {\bibinfo {volume} {85}},\ \bibinfo
  {pages} {054503} (\bibinfo {year} {2012})},\ \Eprint
  {https://arxiv.org/abs/1111.1710} {arXiv:1111.1710 [hep-lat]} \BibitemShut
  {NoStop}%
\bibitem [{\citenamefont {Borsanyi}\ \emph
  {et~al.}(2010{\natexlab{b}})\citenamefont {Borsanyi}, \citenamefont {Fodor},
  \citenamefont {Hoelbling}, \citenamefont {Katz}, \citenamefont {Krieg},
  \citenamefont {Ratti},\ and\ \citenamefont {Szabo}}]{Borsanyi:2010bp}%
  \BibitemOpen
  \bibfield  {author} {\bibinfo {author} {\bibfnamefont {S.}~\bibnamefont
  {Borsanyi}}, \bibinfo {author} {\bibfnamefont {Z.}~\bibnamefont {Fodor}},
  \bibinfo {author} {\bibfnamefont {C.}~\bibnamefont {Hoelbling}}, \bibinfo
  {author} {\bibfnamefont {S.~D.}\ \bibnamefont {Katz}}, \bibinfo {author}
  {\bibfnamefont {S.}~\bibnamefont {Krieg}}, \bibinfo {author} {\bibfnamefont
  {C.}~\bibnamefont {Ratti}},\ and\ \bibinfo {author} {\bibfnamefont {K.~K.}\
  \bibnamefont {Szabo}} (\bibinfo {collaboration} {Wuppertal-Budapest}),\
  }\href {https://doi.org/10.1007/JHEP09(2010)073} {\bibfield  {journal}
  {\bibinfo  {journal} {JHEP}\ }\textbf {\bibinfo {volume} {09}},\ \bibinfo
  {pages} {073} (\bibinfo {year} {2010}{\natexlab{b}})},\ \Eprint
  {https://arxiv.org/abs/1005.3508} {arXiv:1005.3508 [hep-lat]} \BibitemShut
  {NoStop}%
\bibitem [{\citenamefont {Borsanyi}\ \emph {et~al.}(2014)\citenamefont
  {Borsanyi}, \citenamefont {Fodor}, \citenamefont {Hoelbling}, \citenamefont
  {Katz}, \citenamefont {Krieg},\ and\ \citenamefont
  {Szabo}}]{Borsanyi:2013bia}%
  \BibitemOpen
  \bibfield  {author} {\bibinfo {author} {\bibfnamefont {S.}~\bibnamefont
  {Borsanyi}}, \bibinfo {author} {\bibfnamefont {Z.}~\bibnamefont {Fodor}},
  \bibinfo {author} {\bibfnamefont {C.}~\bibnamefont {Hoelbling}}, \bibinfo
  {author} {\bibfnamefont {S.~D.}\ \bibnamefont {Katz}}, \bibinfo {author}
  {\bibfnamefont {S.}~\bibnamefont {Krieg}},\ and\ \bibinfo {author}
  {\bibfnamefont {K.~K.}\ \bibnamefont {Szabo}},\ }\href
  {https://doi.org/10.1016/j.physletb.2014.01.007} {\bibfield  {journal}
  {\bibinfo  {journal} {Phys. Lett. B}\ }\textbf {\bibinfo {volume} {730}},\
  \bibinfo {pages} {99} (\bibinfo {year} {2014})},\ \Eprint
  {https://arxiv.org/abs/1309.5258} {arXiv:1309.5258 [hep-lat]} \BibitemShut
  {NoStop}%
\bibitem [{\citenamefont {Cheng}\ \emph {et~al.}(2010)\citenamefont {Cheng}
  \emph {et~al.}}]{Cheng:2009zi}%
  \BibitemOpen
  \bibfield  {author} {\bibinfo {author} {\bibfnamefont {M.}~\bibnamefont
  {Cheng}} \emph {et~al.},\ }\href {https://doi.org/10.1103/PhysRevD.81.054504}
  {\bibfield  {journal} {\bibinfo  {journal} {Phys. Rev. D}\ }\textbf {\bibinfo
  {volume} {81}},\ \bibinfo {pages} {054504} (\bibinfo {year} {2010})},\
  \Eprint {https://arxiv.org/abs/0911.2215} {arXiv:0911.2215 [hep-lat]}
  \BibitemShut {NoStop}%
\bibitem [{\citenamefont {Alford}, \citenamefont {Rajagopal},\ and\
  \citenamefont {Wilczek}(1998)}]{98Alford.Rajagopal.ea247-256PLB}%
  \BibitemOpen
  \bibfield  {author} {\bibinfo {author} {\bibfnamefont {M.~G.}\ \bibnamefont
  {Alford}}, \bibinfo {author} {\bibfnamefont {K.}~\bibnamefont {Rajagopal}},\
  and\ \bibinfo {author} {\bibfnamefont {F.}~\bibnamefont {Wilczek}},\ }\href
  {https://doi.org/10.1016/S0370-2693(98)00051-3} {\bibfield  {journal}
  {\bibinfo  {journal} {Phys. Lett. B}\ }\textbf {\bibinfo {volume} {422}},\
  \bibinfo {pages} {247} (\bibinfo {year} {1998})},\ \Eprint
  {https://arxiv.org/abs/hep-ph/9711395} {arXiv:hep-ph/9711395 [hep-ph]}
  \BibitemShut {NoStop}%
\bibitem [{\citenamefont {Halasz}\ \emph {et~al.}(1998)\citenamefont {Halasz},
  \citenamefont {Jackson}, \citenamefont {Shrock}, \citenamefont {Stephanov},\
  and\ \citenamefont {Verbaarschot}}]{98Halasz.Jackson.ea96007-96007PRD}%
  \BibitemOpen
  \bibfield  {author} {\bibinfo {author} {\bibfnamefont {A.~M.}\ \bibnamefont
  {Halasz}}, \bibinfo {author} {\bibfnamefont {A.~D.}\ \bibnamefont {Jackson}},
  \bibinfo {author} {\bibfnamefont {R.~E.}\ \bibnamefont {Shrock}}, \bibinfo
  {author} {\bibfnamefont {M.~A.}\ \bibnamefont {Stephanov}},\ and\ \bibinfo
  {author} {\bibfnamefont {J.~J.~M.}\ \bibnamefont {Verbaarschot}},\ }\href
  {https://doi.org/10.1103/PhysRevD.58.096007} {\bibfield  {journal} {\bibinfo
  {journal} {Phys. Rev. D}\ }\textbf {\bibinfo {volume} {58}},\ \bibinfo
  {pages} {096007} (\bibinfo {year} {1998})},\ \Eprint
  {https://arxiv.org/abs/hep-ph/9804290} {arXiv:hep-ph/9804290 [hep-ph]}
  \BibitemShut {NoStop}%
\bibitem [{\citenamefont {Costa}, \citenamefont {Ruivo},\ and\ \citenamefont
  {de~Sousa}(2008)}]{08Costa.Ruivo.ea96001-96001PRD}%
  \BibitemOpen
  \bibfield  {author} {\bibinfo {author} {\bibfnamefont {P.}~\bibnamefont
  {Costa}}, \bibinfo {author} {\bibfnamefont {M.~C.}\ \bibnamefont {Ruivo}},\
  and\ \bibinfo {author} {\bibfnamefont {C.~A.}\ \bibnamefont {de~Sousa}},\
  }\href {https://doi.org/10.1103/PhysRevD.77.096001} {\bibfield  {journal}
  {\bibinfo  {journal} {Phys. Rev. D}\ }\textbf {\bibinfo {volume} {77}},\
  \bibinfo {pages} {096001} (\bibinfo {year} {2008})},\ \Eprint
  {https://arxiv.org/abs/0801.3417} {arXiv:0801.3417 [hep-ph]} \BibitemShut
  {NoStop}%
\bibitem [{\citenamefont {Ferreira}, \citenamefont {Costa},\ and\ \citenamefont
  {Providência}(2018)}]{18Ferreira.Costa.ea34006-34006PRD}%
  \BibitemOpen
  \bibfield  {author} {\bibinfo {author} {\bibfnamefont {M.}~\bibnamefont
  {Ferreira}}, \bibinfo {author} {\bibfnamefont {P.}~\bibnamefont {Costa}},\
  and\ \bibinfo {author} {\bibfnamefont {C.}~\bibnamefont {Providência}},\
  }\href {https://doi.org/10.1103/PhysRevD.98.034006} {\bibfield  {journal}
  {\bibinfo  {journal} {Phys. Rev. D}\ }\textbf {\bibinfo {volume} {98}},\
  \bibinfo {pages} {034006} (\bibinfo {year} {2018})},\ \Eprint
  {https://arxiv.org/abs/1806.05757} {arXiv:1806.05757 [hep-ph]} \BibitemShut
  {NoStop}%
\bibitem [{\citenamefont {Qin}\ \emph {et~al.}(2011)\citenamefont {Qin},
  \citenamefont {Chang}, \citenamefont {Chen}, \citenamefont {Liu},\ and\
  \citenamefont {Roberts}}]{11Qin.Chang.ea172301-172301PRL}%
  \BibitemOpen
  \bibfield  {author} {\bibinfo {author} {\bibfnamefont {S.-x.}\ \bibnamefont
  {Qin}}, \bibinfo {author} {\bibfnamefont {L.}~\bibnamefont {Chang}}, \bibinfo
  {author} {\bibfnamefont {H.}~\bibnamefont {Chen}}, \bibinfo {author}
  {\bibfnamefont {Y.-x.}\ \bibnamefont {Liu}},\ and\ \bibinfo {author}
  {\bibfnamefont {C.~D.}\ \bibnamefont {Roberts}},\ }\href
  {https://doi.org/10.1103/PhysRevLett.106.172301} {\bibfield  {journal}
  {\bibinfo  {journal} {Phys. Rev. Lett.}\ }\textbf {\bibinfo {volume} {106}},\
  \bibinfo {pages} {172301} (\bibinfo {year} {2011})},\ \Eprint
  {https://arxiv.org/abs/1011.2876} {arXiv:1011.2876 [nucl-th]} \BibitemShut
  {NoStop}%
\bibitem [{\citenamefont {Costa}\ \emph {et~al.}(2014)\citenamefont {Costa},
  \citenamefont {Ferreira}, \citenamefont {Hansen}, \citenamefont {Menezes},\
  and\ \citenamefont {Providência}}]{14Costa.Ferreira.ea56013-56013PRD}%
  \BibitemOpen
  \bibfield  {author} {\bibinfo {author} {\bibfnamefont {P.}~\bibnamefont
  {Costa}}, \bibinfo {author} {\bibfnamefont {M.}~\bibnamefont {Ferreira}},
  \bibinfo {author} {\bibfnamefont {H.}~\bibnamefont {Hansen}}, \bibinfo
  {author} {\bibfnamefont {D.~P.}\ \bibnamefont {Menezes}},\ and\ \bibinfo
  {author} {\bibfnamefont {C.}~\bibnamefont {Providência}},\ }\href
  {https://doi.org/10.1103/PhysRevD.89.056013} {\bibfield  {journal} {\bibinfo
  {journal} {Phys. Rev. D}\ }\textbf {\bibinfo {volume} {89}},\ \bibinfo
  {pages} {056013} (\bibinfo {year} {2014})},\ \Eprint
  {https://arxiv.org/abs/1307.7894} {arXiv:1307.7894 [hep-ph]} \BibitemShut
  {NoStop}%
\bibitem [{\citenamefont {Ruggieri}\ \emph {et~al.}(2014)\citenamefont
  {Ruggieri}, \citenamefont {Oliva}, \citenamefont {Castorina}, \citenamefont
  {Gatto},\ and\ \citenamefont {Greco}}]{14Ruggieri.Oliva.ea255-260PLB}%
  \BibitemOpen
  \bibfield  {author} {\bibinfo {author} {\bibfnamefont {M.}~\bibnamefont
  {Ruggieri}}, \bibinfo {author} {\bibfnamefont {L.}~\bibnamefont {Oliva}},
  \bibinfo {author} {\bibfnamefont {P.}~\bibnamefont {Castorina}}, \bibinfo
  {author} {\bibfnamefont {R.}~\bibnamefont {Gatto}},\ and\ \bibinfo {author}
  {\bibfnamefont {V.}~\bibnamefont {Greco}},\ }\href
  {https://doi.org/10.1016/j.physletb.2014.05.073} {\bibfield  {journal}
  {\bibinfo  {journal} {Phys. Lett. B}\ }\textbf {\bibinfo {volume} {734}},\
  \bibinfo {pages} {255} (\bibinfo {year} {2014})},\ \Eprint
  {https://arxiv.org/abs/1402.0737} {arXiv:1402.0737 [hep-ph]} \BibitemShut
  {NoStop}%
\bibitem [{\citenamefont {Costa}\ \emph {et~al.}(2015)\citenamefont {Costa},
  \citenamefont {Ferreira}, \citenamefont {Menezes}, \citenamefont {Moreira},\
  and\ \citenamefont {Providência}}]{15Costa.Ferreira.ea36012-36012PRD}%
  \BibitemOpen
  \bibfield  {author} {\bibinfo {author} {\bibfnamefont {P.}~\bibnamefont
  {Costa}}, \bibinfo {author} {\bibfnamefont {M.}~\bibnamefont {Ferreira}},
  \bibinfo {author} {\bibfnamefont {D.~P.}\ \bibnamefont {Menezes}}, \bibinfo
  {author} {\bibfnamefont {J.}~\bibnamefont {Moreira}},\ and\ \bibinfo {author}
  {\bibfnamefont {C.}~\bibnamefont {Providência}},\ }\href
  {https://doi.org/10.1103/PhysRevD.92.036012} {\bibfield  {journal} {\bibinfo
  {journal} {Phys. Rev. D}\ }\textbf {\bibinfo {volume} {92}},\ \bibinfo
  {pages} {036012} (\bibinfo {year} {2015})},\ \Eprint
  {https://arxiv.org/abs/1508.07870} {arXiv:1508.07870 [hep-ph]} \BibitemShut
  {NoStop}%
\bibitem [{\citenamefont {Lu}\ \emph {et~al.}(2015)\citenamefont {Lu},
  \citenamefont {Du}, \citenamefont {Cui},\ and\ \citenamefont
  {Zong}}]{15Lu.Du.ea495-495EPJC}%
  \BibitemOpen
  \bibfield  {author} {\bibinfo {author} {\bibfnamefont {Y.}~\bibnamefont
  {Lu}}, \bibinfo {author} {\bibfnamefont {Y.-L.}\ \bibnamefont {Du}}, \bibinfo
  {author} {\bibfnamefont {Z.-F.}\ \bibnamefont {Cui}},\ and\ \bibinfo {author}
  {\bibfnamefont {H.-S.}\ \bibnamefont {Zong}},\ }\href
  {https://doi.org/10.1140/epjc/s10052-015-3720-2} {\bibfield  {journal}
  {\bibinfo  {journal} {Eur. Phys. J. C}\ }\textbf {\bibinfo {volume} {75}},\
  \bibinfo {pages} {495} (\bibinfo {year} {2015})},\ \Eprint
  {https://arxiv.org/abs/1508.00651} {arXiv:1508.00651 [hep-ph]} \BibitemShut
  {NoStop}%
\bibitem [{\citenamefont {Ruggieri}(2011)}]{11Ruggieri14011-14011PRD}%
  \BibitemOpen
  \bibfield  {author} {\bibinfo {author} {\bibfnamefont {M.}~\bibnamefont
  {Ruggieri}},\ }\href {https://doi.org/10.1103/PhysRevD.84.014011} {\bibfield
  {journal} {\bibinfo  {journal} {Phys. Rev. D}\ }\textbf {\bibinfo {volume}
  {84}},\ \bibinfo {pages} {014011} (\bibinfo {year} {2011})},\ \Eprint
  {https://arxiv.org/abs/1103.6186} {arXiv:1103.6186 [hep-ph]} \BibitemShut
  {NoStop}%
\bibitem [{\citenamefont {Jeon}\ and\ \citenamefont
  {Koch}(2000)}]{00Jeon.Koch2076-2079PRL}%
  \BibitemOpen
  \bibfield  {author} {\bibinfo {author} {\bibfnamefont {S.}~\bibnamefont
  {Jeon}}\ and\ \bibinfo {author} {\bibfnamefont {V.}~\bibnamefont {Koch}},\
  }\href {https://doi.org/10.1103/PhysRevLett.85.2076} {\bibfield  {journal}
  {\bibinfo  {journal} {Phys. Rev. Lett.}\ }\textbf {\bibinfo {volume} {85}},\
  \bibinfo {pages} {2076} (\bibinfo {year} {2000})},\ \Eprint
  {https://arxiv.org/abs/hep-ph/0003168} {arXiv:hep-ph/0003168 [hep-ph]}
  \BibitemShut {NoStop}%
\bibitem [{\citenamefont {Asakawa}, \citenamefont {Heinz},\ and\ \citenamefont
  {Muller}(2000)}]{00Asakawa.Heinz.ea2072-2075PRL}%
  \BibitemOpen
  \bibfield  {author} {\bibinfo {author} {\bibfnamefont {M.}~\bibnamefont
  {Asakawa}}, \bibinfo {author} {\bibfnamefont {U.~W.}\ \bibnamefont {Heinz}},\
  and\ \bibinfo {author} {\bibfnamefont {B.}~\bibnamefont {Muller}},\ }\href
  {https://doi.org/10.1103/PhysRevLett.85.2072} {\bibfield  {journal} {\bibinfo
   {journal} {Phys. Rev. Lett.}\ }\textbf {\bibinfo {volume} {85}},\ \bibinfo
  {pages} {2072} (\bibinfo {year} {2000})},\ \Eprint
  {https://arxiv.org/abs/hep-ph/0003169} {arXiv:hep-ph/0003169 [hep-ph]}
  \BibitemShut {NoStop}%
\bibitem [{\citenamefont {Hatta}\ and\ \citenamefont
  {Ikeda}(2003)}]{03Hatta.Ikeda14028-14028PRD}%
  \BibitemOpen
  \bibfield  {author} {\bibinfo {author} {\bibfnamefont {Y.}~\bibnamefont
  {Hatta}}\ and\ \bibinfo {author} {\bibfnamefont {T.}~\bibnamefont {Ikeda}},\
  }\href {https://doi.org/10.1103/PhysRevD.67.014028} {\bibfield  {journal}
  {\bibinfo  {journal} {Phys. Rev. D}\ }\textbf {\bibinfo {volume} {67}},\
  \bibinfo {pages} {014028} (\bibinfo {year} {2003})},\ \Eprint
  {https://arxiv.org/abs/hep-ph/0210284} {arXiv:hep-ph/0210284 [hep-ph]}
  \BibitemShut {NoStop}%
\bibitem [{\citenamefont {Hatta}\ and\ \citenamefont
  {Stephanov}(2003)}]{03Hatta.Stephanov102003-102003PRL}%
  \BibitemOpen
  \bibfield  {author} {\bibinfo {author} {\bibfnamefont {Y.}~\bibnamefont
  {Hatta}}\ and\ \bibinfo {author} {\bibfnamefont {M.~A.}\ \bibnamefont
  {Stephanov}},\ }\href {https://doi.org/10.1103/PhysRevLett.91.102003,
  10.1103/PhysRevLett.91.129901} {\bibfield  {journal} {\bibinfo  {journal}
  {Phys. Rev. Lett.}\ }\textbf {\bibinfo {volume} {91}},\ \bibinfo {pages}
  {102003} (\bibinfo {year} {2003})},\ \bibinfo {note} {[Erratum: Phys. Rev.
  Lett.91,129901(2003)]},\ \Eprint {https://arxiv.org/abs/hep-ph/0302002}
  {arXiv:hep-ph/0302002 [hep-ph]} \BibitemShut {NoStop}%
\bibitem [{\citenamefont {Wang}, \citenamefont {Sun},\ and\ \citenamefont
  {Zong}(2013)}]{13Wang.Sun.ea1350064-1350064MPLA}%
  \BibitemOpen
  \bibfield  {author} {\bibinfo {author} {\bibfnamefont {B.}~\bibnamefont
  {Wang}}, \bibinfo {author} {\bibfnamefont {W.-M.}\ \bibnamefont {Sun}},\ and\
  \bibinfo {author} {\bibfnamefont {H.-S.}\ \bibnamefont {Zong}},\ }\href
  {https://doi.org/10.1142/S0217732313500648} {\bibfield  {journal} {\bibinfo
  {journal} {Mod. Phys. Lett. A}\ }\textbf {\bibinfo {volume} {28}},\ \bibinfo
  {pages} {1350064} (\bibinfo {year} {2013})}\BibitemShut {NoStop}%
\bibitem [{\citenamefont {Min}\ \emph {et~al.}(2008)\citenamefont {Min},
  \citenamefont {Yu}, \citenamefont {Wei-Min},\ and\ \citenamefont
  {Hong-Shi}}]{08Min.Yu.ea76008-76008PRD}%
  \BibitemOpen
  \bibfield  {author} {\bibinfo {author} {\bibfnamefont {H.}~\bibnamefont
  {Min}}, \bibinfo {author} {\bibfnamefont {J.}~\bibnamefont {Yu}}, \bibinfo
  {author} {\bibfnamefont {S.}~\bibnamefont {Wei-Min}},\ and\ \bibinfo {author}
  {\bibfnamefont {Z.}~\bibnamefont {Hong-Shi}},\ }\href
  {https://doi.org/10.1103/PhysRevD.77.076008} {\bibfield  {journal} {\bibinfo
  {journal} {Phys. Rev. D}\ }\textbf {\bibinfo {volume} {77}},\ \bibinfo
  {pages} {076008} (\bibinfo {year} {2008})}\BibitemShut {NoStop}%
\bibitem [{\citenamefont {Chakraborty}, \citenamefont {Mustafa},\ and\
  \citenamefont {Thoma}(2003)}]{03Chakraborty.Mustafa.ea114004-114004PRD}%
  \BibitemOpen
  \bibfield  {author} {\bibinfo {author} {\bibfnamefont {P.}~\bibnamefont
  {Chakraborty}}, \bibinfo {author} {\bibfnamefont {M.~G.}\ \bibnamefont
  {Mustafa}},\ and\ \bibinfo {author} {\bibfnamefont {M.~H.}\ \bibnamefont
  {Thoma}},\ }\href {https://doi.org/10.1103/PhysRevD.67.114004} {\bibfield
  {journal} {\bibinfo  {journal} {Phys. Rev. D}\ }\textbf {\bibinfo {volume}
  {67}},\ \bibinfo {pages} {114004} (\bibinfo {year} {2003})},\ \Eprint
  {https://arxiv.org/abs/hep-ph/0210159} {arXiv:hep-ph/0210159 [hep-ph]}
  \BibitemShut {NoStop}%
\bibitem [{\citenamefont {Zhao}\ \emph {et~al.}(2008)\citenamefont {Zhao},
  \citenamefont {Chang}, \citenamefont {Yuan},\ and\ \citenamefont
  {Liu}}]{08Zhao.Chang.ea483-492EPJC}%
  \BibitemOpen
  \bibfield  {author} {\bibinfo {author} {\bibfnamefont {Y.}~\bibnamefont
  {Zhao}}, \bibinfo {author} {\bibfnamefont {L.}~\bibnamefont {Chang}},
  \bibinfo {author} {\bibfnamefont {W.}~\bibnamefont {Yuan}},\ and\ \bibinfo
  {author} {\bibfnamefont {Y.-x.}\ \bibnamefont {Liu}},\ }\href
  {https://doi.org/10.1140/epjc/s10052-008-0673-8} {\bibfield  {journal}
  {\bibinfo  {journal} {Eur. Phys. J. C}\ }\textbf {\bibinfo {volume} {56}},\
  \bibinfo {pages} {483} (\bibinfo {year} {2008})},\ \Eprint
  {https://arxiv.org/abs/hep-ph/0610358} {arXiv:hep-ph/0610358 [hep-ph]}
  \BibitemShut {NoStop}%
\bibitem [{\citenamefont {Bernard}\ \emph {et~al.}(2005)\citenamefont
  {Bernard}, \citenamefont {Burch}, \citenamefont {Gregory}, \citenamefont
  {Toussaint}, \citenamefont {DeTar}, \citenamefont {Osborn}, \citenamefont
  {Gottlieb}, \citenamefont {Heller},\ and\ \citenamefont
  {Sugar}}]{05Bernard.Burch.ea34504-34504PRD}%
  \BibitemOpen
  \bibfield  {author} {\bibinfo {author} {\bibfnamefont {C.}~\bibnamefont
  {Bernard}}, \bibinfo {author} {\bibfnamefont {T.}~\bibnamefont {Burch}},
  \bibinfo {author} {\bibfnamefont {E.~B.}\ \bibnamefont {Gregory}}, \bibinfo
  {author} {\bibfnamefont {D.}~\bibnamefont {Toussaint}}, \bibinfo {author}
  {\bibfnamefont {C.~E.}\ \bibnamefont {DeTar}}, \bibinfo {author}
  {\bibfnamefont {J.}~\bibnamefont {Osborn}}, \bibinfo {author} {\bibfnamefont
  {S.}~\bibnamefont {Gottlieb}}, \bibinfo {author} {\bibfnamefont {U.~M.}\
  \bibnamefont {Heller}},\ and\ \bibinfo {author} {\bibfnamefont
  {R.}~\bibnamefont {Sugar}} (\bibinfo {collaboration} {MILC}),\ }\href
  {https://doi.org/10.1103/PhysRevD.71.034504} {\bibfield  {journal} {\bibinfo
  {journal} {Phys. Rev. D}\ }\textbf {\bibinfo {volume} {71}},\ \bibinfo
  {pages} {034504} (\bibinfo {year} {2005})},\ \Eprint
  {https://arxiv.org/abs/hep-lat/0405029} {arXiv:hep-lat/0405029 [hep-lat]}
  \BibitemShut {NoStop}%
\bibitem [{\citenamefont {Aoki}\ \emph {et~al.}(2006)\citenamefont {Aoki},
  \citenamefont {Fodor}, \citenamefont {Katz},\ and\ \citenamefont
  {Szabo}}]{06Aoki.Fodor.ea46-54PLB}%
  \BibitemOpen
  \bibfield  {author} {\bibinfo {author} {\bibfnamefont {Y.}~\bibnamefont
  {Aoki}}, \bibinfo {author} {\bibfnamefont {Z.}~\bibnamefont {Fodor}},
  \bibinfo {author} {\bibfnamefont {S.~D.}\ \bibnamefont {Katz}},\ and\
  \bibinfo {author} {\bibfnamefont {K.~K.}\ \bibnamefont {Szabo}},\ }\href
  {https://doi.org/10.1016/j.physletb.2006.10.021} {\bibfield  {journal}
  {\bibinfo  {journal} {Phys. Lett. B}\ }\textbf {\bibinfo {volume} {643}},\
  \bibinfo {pages} {46} (\bibinfo {year} {2006})},\ \Eprint
  {https://arxiv.org/abs/hep-lat/0609068} {arXiv:hep-lat/0609068 [hep-lat]}
  \BibitemShut {NoStop}%
\bibitem [{\citenamefont {Cheng}\ \emph {et~al.}(2007)\citenamefont {Cheng}
  \emph {et~al.}}]{07Cheng.others34506-34506PRD}%
  \BibitemOpen
  \bibfield  {author} {\bibinfo {author} {\bibfnamefont {M.}~\bibnamefont
  {Cheng}} \emph {et~al.},\ }\href {https://doi.org/10.1103/PhysRevD.75.034506}
  {\bibfield  {journal} {\bibinfo  {journal} {Phys. Rev. D}\ }\textbf {\bibinfo
  {volume} {75}},\ \bibinfo {pages} {034506} (\bibinfo {year} {2007})},\
  \Eprint {https://arxiv.org/abs/hep-lat/0612001} {arXiv:hep-lat/0612001
  [hep-lat]} \BibitemShut {NoStop}%
\bibitem [{\citenamefont {Harada}, \citenamefont {Sasaki},\ and\ \citenamefont
  {Takemoto}(2010)}]{10Harada.Sasaki.ea16009-16009PRD}%
  \BibitemOpen
  \bibfield  {author} {\bibinfo {author} {\bibfnamefont {M.}~\bibnamefont
  {Harada}}, \bibinfo {author} {\bibfnamefont {C.}~\bibnamefont {Sasaki}},\
  and\ \bibinfo {author} {\bibfnamefont {S.}~\bibnamefont {Takemoto}},\ }\href
  {https://doi.org/10.1103/PhysRevD.81.016009} {\bibfield  {journal} {\bibinfo
  {journal} {Phys. Rev. D}\ }\textbf {\bibinfo {volume} {81}},\ \bibinfo
  {pages} {016009} (\bibinfo {year} {2010})},\ \Eprint
  {https://arxiv.org/abs/0908.1361} {arXiv:0908.1361 [hep-ph]} \BibitemShut
  {NoStop}%
\bibitem [{\citenamefont {Ding}\ \emph {et~al.}(2015)\citenamefont {Ding},
  \citenamefont {Mukherjee}, \citenamefont {Ohno}, \citenamefont {Petreczky},\
  and\ \citenamefont {Schadler}}]{15Ding.Mukherjee.ea74043-74043PRD}%
  \BibitemOpen
  \bibfield  {author} {\bibinfo {author} {\bibfnamefont {H.~T.}\ \bibnamefont
  {Ding}}, \bibinfo {author} {\bibfnamefont {S.}~\bibnamefont {Mukherjee}},
  \bibinfo {author} {\bibfnamefont {H.}~\bibnamefont {Ohno}}, \bibinfo {author}
  {\bibfnamefont {P.}~\bibnamefont {Petreczky}},\ and\ \bibinfo {author}
  {\bibfnamefont {H.~P.}\ \bibnamefont {Schadler}},\ }\href
  {https://doi.org/10.1103/PhysRevD.92.074043} {\bibfield  {journal} {\bibinfo
  {journal} {Phys. Rev. D}\ }\textbf {\bibinfo {volume} {92}},\ \bibinfo
  {pages} {074043} (\bibinfo {year} {2015})},\ \Eprint
  {https://arxiv.org/abs/1507.06637} {arXiv:1507.06637 [hep-lat]} \BibitemShut
  {NoStop}%
\bibitem [{\citenamefont {Fujii}(2003)}]{03Fujii94018-94018PRD}%
  \BibitemOpen
  \bibfield  {author} {\bibinfo {author} {\bibfnamefont {H.}~\bibnamefont
  {Fujii}},\ }\href {https://doi.org/10.1103/PhysRevD.67.094018} {\bibfield
  {journal} {\bibinfo  {journal} {Phys. Rev. D}\ }\textbf {\bibinfo {volume}
  {67}},\ \bibinfo {pages} {094018} (\bibinfo {year} {2003})}\BibitemShut
  {NoStop}%
\bibitem [{\citenamefont {Fujii}\ and\ \citenamefont
  {Ohtani}(2004)}]{04Fujii.Ohtani14016-14016PRD}%
  \BibitemOpen
  \bibfield  {author} {\bibinfo {author} {\bibfnamefont {H.}~\bibnamefont
  {Fujii}}\ and\ \bibinfo {author} {\bibfnamefont {M.}~\bibnamefont {Ohtani}},\
  }\href {https://doi.org/10.1103/PhysRevD.70.014016} {\bibfield  {journal}
  {\bibinfo  {journal} {Phys. Rev. D}\ }\textbf {\bibinfo {volume} {70}},\
  \bibinfo {pages} {014016} (\bibinfo {year} {2004})},\ \Eprint
  {https://arxiv.org/abs/hep-ph/0402263} {arXiv:hep-ph/0402263 [hep-ph]}
  \BibitemShut {NoStop}%
\bibitem [{\citenamefont {Kunihiro}(1991)}]{91Kunihiro395-402PLB}%
  \BibitemOpen
  \bibfield  {author} {\bibinfo {author} {\bibfnamefont {T.}~\bibnamefont
  {Kunihiro}},\ }\href {https://doi.org/10.1016/0370-2693(91)90107-2}
  {\bibfield  {journal} {\bibinfo  {journal} {Phys. Lett. B}\ }\textbf
  {\bibinfo {volume} {271}},\ \bibinfo {pages} {395} (\bibinfo {year}
  {1991})}\BibitemShut {NoStop}%
\bibitem [{\citenamefont {Stephanov}, \citenamefont {Rajagopal},\ and\
  \citenamefont {Shuryak}(1998)}]{98Stephanov.Rajagopal.ea4816-4819PRL}%
  \BibitemOpen
  \bibfield  {author} {\bibinfo {author} {\bibfnamefont {M.~A.}\ \bibnamefont
  {Stephanov}}, \bibinfo {author} {\bibfnamefont {K.}~\bibnamefont
  {Rajagopal}},\ and\ \bibinfo {author} {\bibfnamefont {E.~V.}\ \bibnamefont
  {Shuryak}},\ }\href {https://doi.org/10.1103/PhysRevLett.81.4816} {\bibfield
  {journal} {\bibinfo  {journal} {Phys. Rev. Lett.}\ }\textbf {\bibinfo
  {volume} {81}},\ \bibinfo {pages} {4816} (\bibinfo {year} {1998})},\ \Eprint
  {https://arxiv.org/abs/hep-ph/9806219} {arXiv:hep-ph/9806219 [hep-ph]}
  \BibitemShut {NoStop}%
\bibitem [{\citenamefont {Gavai}\ and\ \citenamefont
  {Gupta}(2001)}]{01Gavai.Gupta74506-74506PRD}%
  \BibitemOpen
  \bibfield  {author} {\bibinfo {author} {\bibfnamefont {R.~V.}\ \bibnamefont
  {Gavai}}\ and\ \bibinfo {author} {\bibfnamefont {S.}~\bibnamefont {Gupta}},\
  }\href {https://doi.org/10.1103/PhysRevD.64.074506} {\bibfield  {journal}
  {\bibinfo  {journal} {Phys. Rev. D}\ }\textbf {\bibinfo {volume} {64}},\
  \bibinfo {pages} {074506} (\bibinfo {year} {2001})},\ \Eprint
  {https://arxiv.org/abs/hep-lat/0103013} {arXiv:hep-lat/0103013 [hep-lat]}
  \BibitemShut {NoStop}%
\bibitem [{\citenamefont {Gavai}, \citenamefont {Gupta},\ and\ \citenamefont
  {Majumdar}(2002)}]{02Gavai.Gupta.ea54506-54506PRD}%
  \BibitemOpen
  \bibfield  {author} {\bibinfo {author} {\bibfnamefont {R.~V.}\ \bibnamefont
  {Gavai}}, \bibinfo {author} {\bibfnamefont {S.}~\bibnamefont {Gupta}},\ and\
  \bibinfo {author} {\bibfnamefont {P.}~\bibnamefont {Majumdar}},\ }\href
  {https://doi.org/10.1103/PhysRevD.65.054506} {\bibfield  {journal} {\bibinfo
  {journal} {Phys. Rev. D}\ }\textbf {\bibinfo {volume} {65}},\ \bibinfo
  {pages} {054506} (\bibinfo {year} {2002})},\ \Eprint
  {https://arxiv.org/abs/hep-lat/0110032} {arXiv:hep-lat/0110032 [hep-lat]}
  \BibitemShut {NoStop}%
\bibitem [{\citenamefont {Cui}, \citenamefont {Takeuchi},\ and\ \citenamefont
  {Wu}(2011)}]{11Cui.Takeuchi.ea76004-76004PRD}%
  \BibitemOpen
  \bibfield  {author} {\bibinfo {author} {\bibfnamefont {L.-X.}\ \bibnamefont
  {Cui}}, \bibinfo {author} {\bibfnamefont {S.}~\bibnamefont {Takeuchi}},\ and\
  \bibinfo {author} {\bibfnamefont {Y.-L.}\ \bibnamefont {Wu}},\ }\href
  {https://doi.org/10.1103/PhysRevD.84.076004} {\bibfield  {journal} {\bibinfo
  {journal} {Phys. Rev. D}\ }\textbf {\bibinfo {volume} {84}},\ \bibinfo
  {pages} {076004} (\bibinfo {year} {2011})},\ \Eprint
  {https://arxiv.org/abs/1107.2738} {arXiv:1107.2738 [hep-ph]} \BibitemShut
  {NoStop}%
\bibitem [{\citenamefont {Jiang}, \citenamefont {Luo},\ and\ \citenamefont
  {Zong}(2011)}]{11Jiang.Luo.ea66-66J}%
  \BibitemOpen
  \bibfield  {author} {\bibinfo {author} {\bibfnamefont {Y.}~\bibnamefont
  {Jiang}}, \bibinfo {author} {\bibfnamefont {L.-J.}\ \bibnamefont {Luo}},\
  and\ \bibinfo {author} {\bibfnamefont {H.-S.}\ \bibnamefont {Zong}},\ }\href
  {https://doi.org/10.1007/JHEP02(2011)066} {\bibfield  {journal} {\bibinfo
  {journal} {JHEP}\ }\textbf {\bibinfo {volume} {02}},\ \bibinfo {pages} {066}
  (\bibinfo {year} {2011})},\ \Eprint {https://arxiv.org/abs/1102.1532}
  {arXiv:1102.1532 [hep-ph]} \BibitemShut {NoStop}%
\bibitem [{\citenamefont {He}\ \emph {et~al.}(2009)\citenamefont {He},
  \citenamefont {Li}, \citenamefont {Sun},\ and\ \citenamefont
  {Zong}}]{09He.Li.ea36001-36001PRD}%
  \BibitemOpen
  \bibfield  {author} {\bibinfo {author} {\bibfnamefont {M.}~\bibnamefont
  {He}}, \bibinfo {author} {\bibfnamefont {J.-F.}\ \bibnamefont {Li}}, \bibinfo
  {author} {\bibfnamefont {W.-M.}\ \bibnamefont {Sun}},\ and\ \bibinfo {author}
  {\bibfnamefont {H.-S.}\ \bibnamefont {Zong}},\ }\href
  {https://doi.org/10.1103/PhysRevD.79.036001} {\bibfield  {journal} {\bibinfo
  {journal} {Phys. Rev. D}\ }\textbf {\bibinfo {volume} {79}},\ \bibinfo
  {pages} {036001} (\bibinfo {year} {2009})},\ \Eprint
  {https://arxiv.org/abs/0811.1835} {arXiv:0811.1835 [hep-ph]} \BibitemShut
  {NoStop}%
\bibitem [{\citenamefont {Ratti}, \citenamefont {Roessner},\ and\ \citenamefont
  {Weise}(2007)}]{07Ratti.Roessner.ea57-60PLB}%
  \BibitemOpen
  \bibfield  {author} {\bibinfo {author} {\bibfnamefont {C.}~\bibnamefont
  {Ratti}}, \bibinfo {author} {\bibfnamefont {S.}~\bibnamefont {Roessner}},\
  and\ \bibinfo {author} {\bibfnamefont {W.}~\bibnamefont {Weise}},\ }\href
  {https://doi.org/10.1016/j.physletb.2007.03.038} {\bibfield  {journal}
  {\bibinfo  {journal} {Phys. Lett. B}\ }\textbf {\bibinfo {volume} {649}},\
  \bibinfo {pages} {57} (\bibinfo {year} {2007})},\ \Eprint
  {https://arxiv.org/abs/hep-ph/0701091} {arXiv:hep-ph/0701091 [hep-ph]}
  \BibitemShut {NoStop}%
\bibitem [{\citenamefont {Brandt}, \citenamefont
  {Endr\ifmmode~\mbox{\H{o}}\else \H{o}\fi{}di},\ and\ \citenamefont
  {Schmalzbauer}(2018)}]{18Brandt.Endrfmmodeboxolseoidi.ea54514-54514PRD}%
  \BibitemOpen
  \bibfield  {author} {\bibinfo {author} {\bibfnamefont {B.~B.}\ \bibnamefont
  {Brandt}}, \bibinfo {author} {\bibfnamefont {G.}~\bibnamefont
  {Endr\ifmmode~\mbox{\H{o}}\else \H{o}\fi{}di}},\ and\ \bibinfo {author}
  {\bibfnamefont {S.}~\bibnamefont {Schmalzbauer}},\ }\href
  {https://doi.org/10.1103/PhysRevD.97.054514} {\bibfield  {journal} {\bibinfo
  {journal} {Phys. Rev. D}\ }\textbf {\bibinfo {volume} {97}},\ \bibinfo
  {pages} {054514} (\bibinfo {year} {2018})},\ \Eprint
  {https://arxiv.org/abs/1712.08190} {arXiv:1712.08190 [hep-lat]} \BibitemShut
  {NoStop}%
\bibitem [{\citenamefont {Adhikari}\ and\ \citenamefont
  {Andersen}(2017)}]{Adhikari-2016vuu}%
  \BibitemOpen
  \bibfield  {author} {\bibinfo {author} {\bibfnamefont {P.}~\bibnamefont
  {Adhikari}}\ and\ \bibinfo {author} {\bibfnamefont {J.~O.}\ \bibnamefont
  {Andersen}},\ }\href {https://doi.org/10.1103/PhysRevD.95.054020} {\bibfield
  {journal} {\bibinfo  {journal} {Phys. Rev. D}\ }\textbf {\bibinfo {volume}
  {95}},\ \bibinfo {pages} {054020} (\bibinfo {year} {2017})},\ \Eprint
  {https://arxiv.org/abs/1610.01647} {arXiv:1610.01647 [hep-th]} \BibitemShut
  {NoStop}%
\bibitem [{\citenamefont {Muroya}\ \emph {et~al.}(2003)\citenamefont {Muroya},
  \citenamefont {Nakamura}, \citenamefont {Nonaka},\ and\ \citenamefont
  {Takaishi}}]{03Muroya.Nakamura.ea615-668PTP}%
  \BibitemOpen
  \bibfield  {author} {\bibinfo {author} {\bibfnamefont {S.}~\bibnamefont
  {Muroya}}, \bibinfo {author} {\bibfnamefont {A.}~\bibnamefont {Nakamura}},
  \bibinfo {author} {\bibfnamefont {C.}~\bibnamefont {Nonaka}},\ and\ \bibinfo
  {author} {\bibfnamefont {T.}~\bibnamefont {Takaishi}},\ }\href
  {https://doi.org/10.1143/PTP.110.615} {\bibfield  {journal} {\bibinfo
  {journal} {Prog. Theor. Phys.}\ }\textbf {\bibinfo {volume} {110}},\ \bibinfo
  {pages} {615} (\bibinfo {year} {2003})},\ \Eprint
  {https://arxiv.org/abs/hep-lat/0306031} {arXiv:hep-lat/0306031 [hep-lat]}
  \BibitemShut {NoStop}%
\bibitem [{\citenamefont {Son}\ and\ \citenamefont
  {Stephanov}(2001)}]{01Son.Stephanov592-595PRL}%
  \BibitemOpen
  \bibfield  {author} {\bibinfo {author} {\bibfnamefont {D.~T.}\ \bibnamefont
  {Son}}\ and\ \bibinfo {author} {\bibfnamefont {M.~A.}\ \bibnamefont
  {Stephanov}},\ }\href {https://doi.org/10.1103/PhysRevLett.86.592} {\bibfield
   {journal} {\bibinfo  {journal} {Phys. Rev. Lett.}\ }\textbf {\bibinfo
  {volume} {86}},\ \bibinfo {pages} {592} (\bibinfo {year} {2001})},\ \Eprint
  {https://arxiv.org/abs/hep-ph/0005225} {arXiv:hep-ph/0005225 [hep-ph]}
  \BibitemShut {NoStop}%
\bibitem [{\citenamefont {Splittorff}, \citenamefont {Son},\ and\ \citenamefont
  {Stephanov}(2001)}]{01Splittorff.Son.ea16003-16003PRD}%
  \BibitemOpen
  \bibfield  {author} {\bibinfo {author} {\bibfnamefont {K.}~\bibnamefont
  {Splittorff}}, \bibinfo {author} {\bibfnamefont {D.~T.}\ \bibnamefont
  {Son}},\ and\ \bibinfo {author} {\bibfnamefont {M.~A.}\ \bibnamefont
  {Stephanov}},\ }\href {https://doi.org/10.1103/PhysRevD.64.016003} {\bibfield
   {journal} {\bibinfo  {journal} {Phys. Rev. D}\ }\textbf {\bibinfo {volume}
  {64}},\ \bibinfo {pages} {016003} (\bibinfo {year} {2001})},\ \Eprint
  {https://arxiv.org/abs/hep-ph/0012274} {arXiv:hep-ph/0012274 [hep-ph]}
  \BibitemShut {NoStop}%
\bibitem [{\citenamefont {Adhikari}, \citenamefont {Andersen},\ and\
  \citenamefont {Kneschke}(2019)}]{Adhikari-2019mdk}%
  \BibitemOpen
  \bibfield  {author} {\bibinfo {author} {\bibfnamefont {P.}~\bibnamefont
  {Adhikari}}, \bibinfo {author} {\bibfnamefont {J.~O.}\ \bibnamefont
  {Andersen}},\ and\ \bibinfo {author} {\bibfnamefont {P.}~\bibnamefont
  {Kneschke}},\ }\href {https://doi.org/10.1140/epjc/s10052-019-7381-4}
  {\bibfield  {journal} {\bibinfo  {journal} {Eur. Phys. J. C}\ }\textbf
  {\bibinfo {volume} {79}},\ \bibinfo {pages} {874} (\bibinfo {year} {2019})},\
  \Eprint {https://arxiv.org/abs/1904.03887} {arXiv:1904.03887 [hep-ph]}
  \BibitemShut {NoStop}%
\bibitem [{\citenamefont {Adhikari}\ and\ \citenamefont
  {Andersen}(2019{\natexlab{a}})}]{Adhikari-2019mlf}%
  \BibitemOpen
  \bibfield  {author} {\bibinfo {author} {\bibfnamefont {P.}~\bibnamefont
  {Adhikari}}\ and\ \bibinfo {author} {\bibfnamefont {J.~O.}\ \bibnamefont
  {Andersen}},\ }\href@noop {} {\  (\bibinfo {year} {2019}{\natexlab{a}})},\
  \Eprint {https://arxiv.org/abs/1909.10575} {arXiv:1909.10575 [hep-ph]}
  \BibitemShut {NoStop}%
\bibitem [{\citenamefont {Kogut}\ and\ \citenamefont
  {Sinclair}(2002{\natexlab{a}})}]{02Kogut.Sinclair34505-34505PRD}%
  \BibitemOpen
  \bibfield  {author} {\bibinfo {author} {\bibfnamefont {J.~B.}\ \bibnamefont
  {Kogut}}\ and\ \bibinfo {author} {\bibfnamefont {D.~K.}\ \bibnamefont
  {Sinclair}},\ }\href {https://doi.org/10.1103/PhysRevD.66.034505} {\bibfield
  {journal} {\bibinfo  {journal} {Phys. Rev. D}\ }\textbf {\bibinfo {volume}
  {66}},\ \bibinfo {pages} {034505} (\bibinfo {year} {2002}{\natexlab{a}})},\
  \Eprint {https://arxiv.org/abs/hep-lat/0202028} {arXiv:hep-lat/0202028
  [hep-lat]} \BibitemShut {NoStop}%
\bibitem [{\citenamefont {Kogut}\ and\ \citenamefont
  {Sinclair}(2002{\natexlab{b}})}]{02Kogut.Sinclair14508-14508PRD}%
  \BibitemOpen
  \bibfield  {author} {\bibinfo {author} {\bibfnamefont {J.~B.}\ \bibnamefont
  {Kogut}}\ and\ \bibinfo {author} {\bibfnamefont {D.~K.}\ \bibnamefont
  {Sinclair}},\ }\href {https://doi.org/10.1103/PhysRevD.66.014508} {\bibfield
  {journal} {\bibinfo  {journal} {Phys. Rev. D}\ }\textbf {\bibinfo {volume}
  {66}},\ \bibinfo {pages} {014508} (\bibinfo {year} {2002}{\natexlab{b}})},\
  \Eprint {https://arxiv.org/abs/hep-lat/0201017} {arXiv:hep-lat/0201017
  [hep-lat]} \BibitemShut {NoStop}%
\bibitem [{\citenamefont {Kogut}\ and\ \citenamefont
  {Sinclair}(2004)}]{04Kogut.Sinclair94501-94501PRD}%
  \BibitemOpen
  \bibfield  {author} {\bibinfo {author} {\bibfnamefont {J.~B.}\ \bibnamefont
  {Kogut}}\ and\ \bibinfo {author} {\bibfnamefont {D.~K.}\ \bibnamefont
  {Sinclair}},\ }\href {https://doi.org/10.1103/PhysRevD.70.094501} {\bibfield
  {journal} {\bibinfo  {journal} {Phys. Rev. D}\ }\textbf {\bibinfo {volume}
  {70}},\ \bibinfo {pages} {094501} (\bibinfo {year} {2004})},\ \Eprint
  {https://arxiv.org/abs/hep-lat/0407027} {arXiv:hep-lat/0407027 [hep-lat]}
  \BibitemShut {NoStop}%
\bibitem [{\citenamefont {Detmold}, \citenamefont {Orginos},\ and\
  \citenamefont {Shi}(2012)}]{12Detmold.Orginos.ea54507-54507PRD}%
  \BibitemOpen
  \bibfield  {author} {\bibinfo {author} {\bibfnamefont {W.}~\bibnamefont
  {Detmold}}, \bibinfo {author} {\bibfnamefont {K.}~\bibnamefont {Orginos}},\
  and\ \bibinfo {author} {\bibfnamefont {Z.}~\bibnamefont {Shi}},\ }\href
  {https://doi.org/10.1103/PhysRevD.86.054507} {\bibfield  {journal} {\bibinfo
  {journal} {Phys. Rev. D}\ }\textbf {\bibinfo {volume} {86}},\ \bibinfo
  {pages} {054507} (\bibinfo {year} {2012})},\ \Eprint
  {https://arxiv.org/abs/1205.4224} {arXiv:1205.4224 [hep-lat]} \BibitemShut
  {NoStop}%
\bibitem [{\citenamefont {He}, \citenamefont {Jin},\ and\ \citenamefont
  {Zhuang}(2005)}]{05He.Jin.ea116001-116001PRD}%
  \BibitemOpen
  \bibfield  {author} {\bibinfo {author} {\bibfnamefont {L.-y.}\ \bibnamefont
  {He}}, \bibinfo {author} {\bibfnamefont {M.}~\bibnamefont {Jin}},\ and\
  \bibinfo {author} {\bibfnamefont {P.-f.}\ \bibnamefont {Zhuang}},\ }\href
  {https://doi.org/10.1103/PhysRevD.71.116001} {\bibfield  {journal} {\bibinfo
  {journal} {Phys. Rev. D}\ }\textbf {\bibinfo {volume} {71}},\ \bibinfo
  {pages} {116001} (\bibinfo {year} {2005})},\ \Eprint
  {https://arxiv.org/abs/hep-ph/0503272} {arXiv:hep-ph/0503272 [hep-ph]}
  \BibitemShut {NoStop}%
\bibitem [{\citenamefont {He}\ and\ \citenamefont
  {Zhuang}(2005)}]{05He.Zhuang93-101PLB}%
  \BibitemOpen
  \bibfield  {author} {\bibinfo {author} {\bibfnamefont {L.}~\bibnamefont
  {He}}\ and\ \bibinfo {author} {\bibfnamefont {P.}~\bibnamefont {Zhuang}},\
  }\href {https://doi.org/10.1016/j.physletb.2005.03.066} {\bibfield  {journal}
  {\bibinfo  {journal} {Phys. Lett. B}\ }\textbf {\bibinfo {volume} {615}},\
  \bibinfo {pages} {93} (\bibinfo {year} {2005})},\ \Eprint
  {https://arxiv.org/abs/hep-ph/0501024} {arXiv:hep-ph/0501024 [hep-ph]}
  \BibitemShut {NoStop}%
\bibitem [{\citenamefont {Warringa}, \citenamefont {Boer},\ and\ \citenamefont
  {Andersen}(2005)}]{05Warringa.Boer.ea14015-14015PRD}%
  \BibitemOpen
  \bibfield  {author} {\bibinfo {author} {\bibfnamefont {H.~J.}\ \bibnamefont
  {Warringa}}, \bibinfo {author} {\bibfnamefont {D.}~\bibnamefont {Boer}},\
  and\ \bibinfo {author} {\bibfnamefont {J.~O.}\ \bibnamefont {Andersen}},\
  }\href {https://doi.org/10.1103/PhysRevD.72.014015} {\bibfield  {journal}
  {\bibinfo  {journal} {Phys. Rev. D}\ }\textbf {\bibinfo {volume} {72}},\
  \bibinfo {pages} {014015} (\bibinfo {year} {2005})},\ \Eprint
  {https://arxiv.org/abs/hep-ph/0504177} {arXiv:hep-ph/0504177 [hep-ph]}
  \BibitemShut {NoStop}%
\bibitem [{\citenamefont {Loewe}\ and\ \citenamefont
  {Villavicencio}(2003)}]{03Loewe.Villavicencio74034-74034PRD}%
  \BibitemOpen
  \bibfield  {author} {\bibinfo {author} {\bibfnamefont {M.}~\bibnamefont
  {Loewe}}\ and\ \bibinfo {author} {\bibfnamefont {C.}~\bibnamefont
  {Villavicencio}},\ }\href {https://doi.org/10.1103/PhysRevD.67.074034}
  {\bibfield  {journal} {\bibinfo  {journal} {Phys. Rev. D}\ }\textbf {\bibinfo
  {volume} {67}},\ \bibinfo {pages} {074034} (\bibinfo {year} {2003})},\
  \Eprint {https://arxiv.org/abs/hep-ph/0212275} {arXiv:hep-ph/0212275
  [hep-ph]} \BibitemShut {NoStop}%
\bibitem [{\citenamefont {Loewe}\ and\ \citenamefont
  {Villavicencio}(2005)}]{05Loewe.Villavicencio94001-94001PRD}%
  \BibitemOpen
  \bibfield  {author} {\bibinfo {author} {\bibfnamefont {M.}~\bibnamefont
  {Loewe}}\ and\ \bibinfo {author} {\bibfnamefont {C.}~\bibnamefont
  {Villavicencio}},\ }\href {https://doi.org/10.1103/PhysRevD.71.094001}
  {\bibfield  {journal} {\bibinfo  {journal} {Phys. Rev. D}\ }\textbf {\bibinfo
  {volume} {71}},\ \bibinfo {pages} {094001} (\bibinfo {year} {2005})},\
  \Eprint {https://arxiv.org/abs/hep-ph/0501261} {arXiv:hep-ph/0501261
  [hep-ph]} \BibitemShut {NoStop}%
\bibitem [{\citenamefont {Adhikari}, \citenamefont {Cohen},\ and\ \citenamefont
  {Sakowitz}(2015)}]{15Adhikari.Cohen.ea45202-45202PRC}%
  \BibitemOpen
  \bibfield  {author} {\bibinfo {author} {\bibfnamefont {P.}~\bibnamefont
  {Adhikari}}, \bibinfo {author} {\bibfnamefont {T.~D.}\ \bibnamefont
  {Cohen}},\ and\ \bibinfo {author} {\bibfnamefont {J.}~\bibnamefont
  {Sakowitz}},\ }\href {https://doi.org/10.1103/PhysRevC.91.045202} {\bibfield
  {journal} {\bibinfo  {journal} {Phys. Rev. C}\ }\textbf {\bibinfo {volume}
  {91}},\ \bibinfo {pages} {045202} (\bibinfo {year} {2015})},\ \Eprint
  {https://arxiv.org/abs/1501.02737} {arXiv:1501.02737 [nucl-th]} \BibitemShut
  {NoStop}%
\bibitem [{\citenamefont {Mammarella}\ and\ \citenamefont
  {Mannarelli}(2015)}]{15Mammarella.Mannarelli85025-85025PRD}%
  \BibitemOpen
  \bibfield  {author} {\bibinfo {author} {\bibfnamefont {A.}~\bibnamefont
  {Mammarella}}\ and\ \bibinfo {author} {\bibfnamefont {M.}~\bibnamefont
  {Mannarelli}},\ }\href {https://doi.org/10.1103/PhysRevD.92.085025}
  {\bibfield  {journal} {\bibinfo  {journal} {Phys. Rev. D}\ }\textbf {\bibinfo
  {volume} {92}},\ \bibinfo {pages} {085025} (\bibinfo {year} {2015})},\
  \Eprint {https://arxiv.org/abs/1507.02934} {arXiv:1507.02934 [hep-ph]}
  \BibitemShut {NoStop}%
\bibitem [{\citenamefont {Carignano}\ \emph {et~al.}(2017)\citenamefont
  {Carignano}, \citenamefont {Lepori}, \citenamefont {Mammarella},
  \citenamefont {Mannarelli},\ and\ \citenamefont
  {Pagliaroli}}]{17Carignano.Lepori.ea35-35EPJA}%
  \BibitemOpen
  \bibfield  {author} {\bibinfo {author} {\bibfnamefont {S.}~\bibnamefont
  {Carignano}}, \bibinfo {author} {\bibfnamefont {L.}~\bibnamefont {Lepori}},
  \bibinfo {author} {\bibfnamefont {A.}~\bibnamefont {Mammarella}}, \bibinfo
  {author} {\bibfnamefont {M.}~\bibnamefont {Mannarelli}},\ and\ \bibinfo
  {author} {\bibfnamefont {G.}~\bibnamefont {Pagliaroli}},\ }\href
  {https://doi.org/10.1140/epja/i2017-12221-x} {\bibfield  {journal} {\bibinfo
  {journal} {Eur. Phys. J. A}\ }\textbf {\bibinfo {volume} {53}},\ \bibinfo
  {pages} {35} (\bibinfo {year} {2017})},\ \Eprint
  {https://arxiv.org/abs/1610.06097} {arXiv:1610.06097 [hep-ph]} \BibitemShut
  {NoStop}%
\bibitem [{\citenamefont {Carignano}, \citenamefont {Mammarella},\ and\
  \citenamefont {Mannarelli}(2016)}]{16Carignano.Mammarella.ea51503-51503PRD}%
  \BibitemOpen
  \bibfield  {author} {\bibinfo {author} {\bibfnamefont {S.}~\bibnamefont
  {Carignano}}, \bibinfo {author} {\bibfnamefont {A.}~\bibnamefont
  {Mammarella}},\ and\ \bibinfo {author} {\bibfnamefont {M.}~\bibnamefont
  {Mannarelli}},\ }\href {https://doi.org/10.1103/PhysRevD.93.051503}
  {\bibfield  {journal} {\bibinfo  {journal} {Phys. Rev. D}\ }\textbf {\bibinfo
  {volume} {93}},\ \bibinfo {pages} {051503} (\bibinfo {year} {2016})},\
  \Eprint {https://arxiv.org/abs/1602.01317} {arXiv:1602.01317 [hep-ph]}
  \BibitemShut {NoStop}%
\bibitem [{\citenamefont {Adhikari}(2019)}]{19Adhikari211-217PLB}%
  \BibitemOpen
  \bibfield  {author} {\bibinfo {author} {\bibfnamefont {P.}~\bibnamefont
  {Adhikari}},\ }\href {https://doi.org/10.1016/j.physletb.2019.01.027}
  {\bibfield  {journal} {\bibinfo  {journal} {Phys. Lett. B}\ }\textbf
  {\bibinfo {volume} {790}},\ \bibinfo {pages} {211} (\bibinfo {year}
  {2019})},\ \Eprint {https://arxiv.org/abs/1810.03663} {arXiv:1810.03663
  [nucl-th]} \BibitemShut {NoStop}%
\bibitem [{\citenamefont {Adhikari}\ and\ \citenamefont
  {Andersen}(2019{\natexlab{b}})}]{Adhikari-2019zaj}%
  \BibitemOpen
  \bibfield  {author} {\bibinfo {author} {\bibfnamefont {P.}~\bibnamefont
  {Adhikari}}\ and\ \bibinfo {author} {\bibfnamefont {J.~O.}\ \bibnamefont
  {Andersen}},\ }\href@noop {} {\  (\bibinfo {year} {2019}{\natexlab{b}})},\
  \Eprint {https://arxiv.org/abs/1909.01131} {arXiv:1909.01131 [hep-ph]}
  \BibitemShut {NoStop}%
\bibitem [{\citenamefont {Brandt}\ and\ \citenamefont
  {Endrodi}(2016)}]{16Brandt.Endrodi39-39P}%
  \BibitemOpen
  \bibfield  {author} {\bibinfo {author} {\bibfnamefont {B.~B.}\ \bibnamefont
  {Brandt}}\ and\ \bibinfo {author} {\bibfnamefont {G.}~\bibnamefont
  {Endrodi}},\ }\bibfield  {booktitle} {\emph {\bibinfo {booktitle}
  {{Proceedings, 34th International Symposium on Lattice Field Theory (Lattice
  2016): Southampton, UK, July 24-30, 2016}}},\ }\href@noop {} {\bibfield
  {journal} {\bibinfo  {journal} {PoS}\ }\textbf {\bibinfo {volume}
  {LATTICE2016}},\ \bibinfo {pages} {039} (\bibinfo {year} {2016})},\ \Eprint
  {https://arxiv.org/abs/1611.06758} {arXiv:1611.06758 [hep-lat]} \BibitemShut
  {NoStop}%
\bibitem [{\citenamefont {Shi}(2013)}]{13Shi12026-12026JPCS}%
  \BibitemOpen
  \bibfield  {author} {\bibinfo {author} {\bibfnamefont {Z.}~\bibnamefont
  {Shi}},\ }\bibfield  {booktitle} {\emph {\bibinfo {booktitle} {{Proceedings,
  Extreme QCD 2012 (XQCD12): Washington, USA, August 21-23, 2012}}},\ }\href
  {https://doi.org/10.1088/1742-6596/432/1/012026} {\bibfield  {journal}
  {\bibinfo  {journal} {J. Phys. Conf. Ser.}\ }\textbf {\bibinfo {volume}
  {432}},\ \bibinfo {pages} {012026} (\bibinfo {year} {2013})},\ \Eprint
  {https://arxiv.org/abs/1211.0481} {arXiv:1211.0481 [hep-lat]} \BibitemShut
  {NoStop}%
\bibitem [{\citenamefont {Scior}, \citenamefont {von Smekal},\ and\
  \citenamefont {Smith}(2018)}]{18Scior.Smekal.ea7042-7042EWC}%
  \BibitemOpen
  \bibfield  {author} {\bibinfo {author} {\bibfnamefont {P.}~\bibnamefont
  {Scior}}, \bibinfo {author} {\bibfnamefont {L.}~\bibnamefont {von Smekal}},\
  and\ \bibinfo {author} {\bibfnamefont {D.}~\bibnamefont {Smith}},\ }\bibfield
   {booktitle} {\emph {\bibinfo {booktitle} {{Proceedings, 35th International
  Symposium on Lattice Field Theory (Lattice 2017): Granada, Spain, June 18-24,
  2017}}},\ }\href {https://doi.org/10.1051/epjconf/201817507042} {\bibfield
  {journal} {\bibinfo  {journal} {EPJ Web Conf.}\ }\textbf {\bibinfo {volume}
  {175}},\ \bibinfo {pages} {07042} (\bibinfo {year} {2018})},\ \Eprint
  {https://arxiv.org/abs/1710.06314} {arXiv:1710.06314 [hep-lat]} \BibitemShut
  {NoStop}%
\bibitem [{\citenamefont {Brandt}, \citenamefont {Endrodi},\ and\ \citenamefont
  {Schmalzbauer}(2018)}]{18Brandt.Endrodi.ea7020-7020EWC}%
  \BibitemOpen
  \bibfield  {author} {\bibinfo {author} {\bibfnamefont {B.~B.}\ \bibnamefont
  {Brandt}}, \bibinfo {author} {\bibfnamefont {G.}~\bibnamefont {Endrodi}},\
  and\ \bibinfo {author} {\bibfnamefont {S.}~\bibnamefont {Schmalzbauer}},\
  }\bibfield  {booktitle} {\emph {\bibinfo {booktitle} {{Proceedings, 35th
  International Symposium on Lattice Field Theory (Lattice 2017): Granada,
  Spain, June 18-24, 2017}}},\ }\href
  {https://doi.org/10.1051/epjconf/201817507020} {\bibfield  {journal}
  {\bibinfo  {journal} {EPJ Web Conf.}\ }\textbf {\bibinfo {volume} {175}},\
  \bibinfo {pages} {07020} (\bibinfo {year} {2018})},\ \Eprint
  {https://arxiv.org/abs/1709.10487} {arXiv:1709.10487 [hep-lat]} \BibitemShut
  {NoStop}%
\bibitem [{\citenamefont {Mao}, \citenamefont {Petropoulos},\ and\
  \citenamefont {Zhao}(2006)}]{06Mao.Petropoulos.ea2187-2198JPG}%
  \BibitemOpen
  \bibfield  {author} {\bibinfo {author} {\bibfnamefont {H.}~\bibnamefont
  {Mao}}, \bibinfo {author} {\bibfnamefont {N.}~\bibnamefont {Petropoulos}},\
  and\ \bibinfo {author} {\bibfnamefont {W.-Q.}\ \bibnamefont {Zhao}},\ }\href
  {https://doi.org/10.1088/0954-3899/32/11/012} {\bibfield  {journal} {\bibinfo
   {journal} {J. Phys. G}\ }\textbf {\bibinfo {volume} {32}},\ \bibinfo {pages}
  {2187} (\bibinfo {year} {2006})},\ \Eprint
  {https://arxiv.org/abs/hep-ph/0606241} {arXiv:hep-ph/0606241 [hep-ph]}
  \BibitemShut {NoStop}%
\bibitem [{\citenamefont {Matsuzaki}(2010)}]{10Matsuzaki16005-16005PRD}%
  \BibitemOpen
  \bibfield  {author} {\bibinfo {author} {\bibfnamefont {M.}~\bibnamefont
  {Matsuzaki}},\ }\href {https://doi.org/10.1103/PhysRevD.82.016005} {\bibfield
   {journal} {\bibinfo  {journal} {Phys. Rev. D}\ }\textbf {\bibinfo {volume}
  {82}},\ \bibinfo {pages} {016005} (\bibinfo {year} {2010})},\ \Eprint
  {https://arxiv.org/abs/0909.5615} {arXiv:0909.5615 [hep-ph]} \BibitemShut
  {NoStop}%
\bibitem [{\citenamefont {Ueda}\ \emph {et~al.}(2013)\citenamefont {Ueda},
  \citenamefont {Nakano}, \citenamefont {Ohnishi}, \citenamefont {Ruggieri},\
  and\ \citenamefont {Sumiyoshi}}]{13Ueda.Nakano.ea74006-74006PRD}%
  \BibitemOpen
  \bibfield  {author} {\bibinfo {author} {\bibfnamefont {H.}~\bibnamefont
  {Ueda}}, \bibinfo {author} {\bibfnamefont {T.~Z.}\ \bibnamefont {Nakano}},
  \bibinfo {author} {\bibfnamefont {A.}~\bibnamefont {Ohnishi}}, \bibinfo
  {author} {\bibfnamefont {M.}~\bibnamefont {Ruggieri}},\ and\ \bibinfo
  {author} {\bibfnamefont {K.}~\bibnamefont {Sumiyoshi}},\ }\href
  {https://doi.org/10.1103/PhysRevD.88.074006} {\bibfield  {journal} {\bibinfo
  {journal} {Phys. Rev. D}\ }\textbf {\bibinfo {volume} {88}},\ \bibinfo
  {pages} {074006} (\bibinfo {year} {2013})},\ \Eprint
  {https://arxiv.org/abs/1304.4331} {arXiv:1304.4331 [nucl-th]} \BibitemShut
  {NoStop}%
\bibitem [{\citenamefont {Stiele}, \citenamefont {Fraga},\ and\ \citenamefont
  {Schaffner-Bielich}(2014)}]{14Stiele.Fraga.ea72-78PLB}%
  \BibitemOpen
  \bibfield  {author} {\bibinfo {author} {\bibfnamefont {R.}~\bibnamefont
  {Stiele}}, \bibinfo {author} {\bibfnamefont {E.~S.}\ \bibnamefont {Fraga}},\
  and\ \bibinfo {author} {\bibfnamefont {J.}~\bibnamefont
  {Schaffner-Bielich}},\ }\href
  {https://doi.org/10.1016/j.physletb.2013.12.053} {\bibfield  {journal}
  {\bibinfo  {journal} {Phys. Lett. B}\ }\textbf {\bibinfo {volume} {729}},\
  \bibinfo {pages} {72} (\bibinfo {year} {2014})},\ \Eprint
  {https://arxiv.org/abs/1307.2851} {arXiv:1307.2851 [hep-ph]} \BibitemShut
  {NoStop}%
\bibitem [{\citenamefont {Nishihara}\ and\ \citenamefont
  {Harada}(2014)}]{14Nishihara.Harada76001-76001PRD}%
  \BibitemOpen
  \bibfield  {author} {\bibinfo {author} {\bibfnamefont {H.}~\bibnamefont
  {Nishihara}}\ and\ \bibinfo {author} {\bibfnamefont {M.}~\bibnamefont
  {Harada}},\ }\href {https://doi.org/10.1103/PhysRevD.89.076001} {\bibfield
  {journal} {\bibinfo  {journal} {Phys. Rev. D}\ }\textbf {\bibinfo {volume}
  {89}},\ \bibinfo {pages} {076001} (\bibinfo {year} {2014})},\ \Eprint
  {https://arxiv.org/abs/1401.2928} {arXiv:1401.2928 [hep-ph]} \BibitemShut
  {NoStop}%
\bibitem [{\citenamefont {Andersen}\ and\ \citenamefont
  {Kyllingstad}(2009)}]{09Andersen.Kyllingstad15003-15003JPG}%
  \BibitemOpen
  \bibfield  {author} {\bibinfo {author} {\bibfnamefont {J.~O.}\ \bibnamefont
  {Andersen}}\ and\ \bibinfo {author} {\bibfnamefont {L.}~\bibnamefont
  {Kyllingstad}},\ }\href {https://doi.org/10.1088/0954-3899/37/1/015003}
  {\bibfield  {journal} {\bibinfo  {journal} {J. Phys. G}\ }\textbf {\bibinfo
  {volume} {37}},\ \bibinfo {pages} {015003} (\bibinfo {year} {2009})},\
  \Eprint {https://arxiv.org/abs/hep-ph/0701033} {arXiv:hep-ph/0701033
  [hep-ph]} \BibitemShut {NoStop}%
\bibitem [{\citenamefont {Xia}, \citenamefont {He},\ and\ \citenamefont
  {Zhuang}(2013)}]{13Xia.He.ea56013-56013PRD}%
  \BibitemOpen
  \bibfield  {author} {\bibinfo {author} {\bibfnamefont {T.}~\bibnamefont
  {Xia}}, \bibinfo {author} {\bibfnamefont {L.}~\bibnamefont {He}},\ and\
  \bibinfo {author} {\bibfnamefont {P.}~\bibnamefont {Zhuang}},\ }\href
  {https://doi.org/10.1103/PhysRevD.88.056013} {\bibfield  {journal} {\bibinfo
  {journal} {Phys. Rev. D}\ }\textbf {\bibinfo {volume} {88}},\ \bibinfo
  {pages} {056013} (\bibinfo {year} {2013})},\ \Eprint
  {https://arxiv.org/abs/1307.4622} {arXiv:1307.4622 [hep-ph]} \BibitemShut
  {NoStop}%
\bibitem [{\citenamefont {Wu}, \citenamefont {Ping},\ and\ \citenamefont
  {Zong}(2017)}]{17Wu.Ping.ea124106-124106CPC}%
  \BibitemOpen
  \bibfield  {author} {\bibinfo {author} {\bibfnamefont {Z.}~\bibnamefont
  {Wu}}, \bibinfo {author} {\bibfnamefont {J.}~\bibnamefont {Ping}},\ and\
  \bibinfo {author} {\bibfnamefont {H.}~\bibnamefont {Zong}},\ }\href
  {https://doi.org/10.1088/1674-1137/41/12/124106} {\bibfield  {journal}
  {\bibinfo  {journal} {Chin. Phys. C}\ }\textbf {\bibinfo {volume} {41}},\
  \bibinfo {pages} {124106} (\bibinfo {year} {2017})}\BibitemShut {NoStop}%
\bibitem [{\citenamefont {Chao}, \citenamefont {Huang},\ and\ \citenamefont
  {Radzhabov}(2018)}]{18Chao.Huang.ea-}%
  \BibitemOpen
  \bibfield  {author} {\bibinfo {author} {\bibfnamefont {J.}~\bibnamefont
  {Chao}}, \bibinfo {author} {\bibfnamefont {M.}~\bibnamefont {Huang}},\ and\
  \bibinfo {author} {\bibfnamefont {A.}~\bibnamefont {Radzhabov}},\ }\href@noop
  {} {\  (\bibinfo {year} {2018})},\ \Eprint {https://arxiv.org/abs/1805.00614}
  {arXiv:1805.00614 [hep-ph]} \BibitemShut {NoStop}%
\bibitem [{\citenamefont {Xiong}, \citenamefont {Jin},\ and\ \citenamefont
  {Li}(2009)}]{09Xiong.Jin.ea125005-125005JPG}%
  \BibitemOpen
  \bibfield  {author} {\bibinfo {author} {\bibfnamefont {J.}~\bibnamefont
  {Xiong}}, \bibinfo {author} {\bibfnamefont {M.}~\bibnamefont {Jin}},\ and\
  \bibinfo {author} {\bibfnamefont {J.}~\bibnamefont {Li}},\ }\href
  {https://doi.org/10.1088/0954-3899/36/12/125005} {\bibfield  {journal}
  {\bibinfo  {journal} {J. Phys. G}\ }\textbf {\bibinfo {volume} {36}},\
  \bibinfo {pages} {125005} (\bibinfo {year} {2009})}\BibitemShut {NoStop}%
\bibitem [{\citenamefont {Abuki}\ \emph {et~al.}(2009)\citenamefont {Abuki},
  \citenamefont {Anglani}, \citenamefont {Gatto}, \citenamefont {Pellicoro},\
  and\ \citenamefont {Ruggieri}}]{Abuki:2008wm}%
  \BibitemOpen
  \bibfield  {author} {\bibinfo {author} {\bibfnamefont {H.}~\bibnamefont
  {Abuki}}, \bibinfo {author} {\bibfnamefont {R.}~\bibnamefont {Anglani}},
  \bibinfo {author} {\bibfnamefont {R.}~\bibnamefont {Gatto}}, \bibinfo
  {author} {\bibfnamefont {M.}~\bibnamefont {Pellicoro}},\ and\ \bibinfo
  {author} {\bibfnamefont {M.}~\bibnamefont {Ruggieri}},\ }\href
  {https://doi.org/10.1103/PhysRevD.79.034032} {\bibfield  {journal} {\bibinfo
  {journal} {Phys. Rev.}\ }\textbf {\bibinfo {volume} {D79}},\ \bibinfo {pages}
  {034032} (\bibinfo {year} {2009})},\ \Eprint
  {https://arxiv.org/abs/0809.2658} {arXiv:0809.2658 [hep-ph]} \BibitemShut
  {NoStop}%
\bibitem [{\citenamefont {Avancini}\ \emph {et~al.}(2019)\citenamefont
  {Avancini}, \citenamefont {Bandyopadhyay}, \citenamefont {Duarte},\ and\
  \citenamefont {Farias}}]{Avancini-2019ego}%
  \BibitemOpen
  \bibfield  {author} {\bibinfo {author} {\bibfnamefont {S.~S.}\ \bibnamefont
  {Avancini}}, \bibinfo {author} {\bibfnamefont {A.}~\bibnamefont
  {Bandyopadhyay}}, \bibinfo {author} {\bibfnamefont {D.~C.}\ \bibnamefont
  {Duarte}},\ and\ \bibinfo {author} {\bibfnamefont {R.~L.~S.}\ \bibnamefont
  {Farias}},\ }\href {https://doi.org/10.1103/PhysRevD.100.116002} {\bibfield
  {journal} {\bibinfo  {journal} {Phys. Rev. D}\ }\textbf {\bibinfo {volume}
  {100}},\ \bibinfo {pages} {116002} (\bibinfo {year} {2019})},\ \Eprint
  {https://arxiv.org/abs/1907.09880} {arXiv:1907.09880 [hep-ph]} \BibitemShut
  {NoStop}%
\bibitem [{\citenamefont {Mannarelli}(2019)}]{Mannarelli-2019hgn}%
  \BibitemOpen
  \bibfield  {author} {\bibinfo {author} {\bibfnamefont {M.}~\bibnamefont
  {Mannarelli}},\ }\href {https://doi.org/10.3390/particles2030025} {\bibfield
  {journal} {\bibinfo  {journal} {Particles}\ }\textbf {\bibinfo {volume}
  {2}},\ \bibinfo {pages} {411} (\bibinfo {year} {2019})},\ \Eprint
  {https://arxiv.org/abs/1908.02042} {arXiv:1908.02042 [hep-ph]} \BibitemShut
  {NoStop}%
\bibitem [{\citenamefont {Mao}(2014)}]{14Mao116006-116006PRD}%
  \BibitemOpen
  \bibfield  {author} {\bibinfo {author} {\bibfnamefont {S.}~\bibnamefont
  {Mao}},\ }\href {https://doi.org/10.1103/PhysRevD.89.116006} {\bibfield
  {journal} {\bibinfo  {journal} {Phys. Rev. D}\ }\textbf {\bibinfo {volume}
  {89}},\ \bibinfo {pages} {116006} (\bibinfo {year} {2014})},\ \Eprint
  {https://arxiv.org/abs/1402.4564} {arXiv:1402.4564 [nucl-th]} \BibitemShut
  {NoStop}%
\bibitem [{\citenamefont {Brandt}\ \emph {et~al.}(2018)\citenamefont {Brandt},
  \citenamefont {Endrodi}, \citenamefont {Fraga}, \citenamefont {Hippert},
  \citenamefont {Schaffner-Bielich},\ and\ \citenamefont
  {Schmalzbauer}}]{18Brandt.Endrodi.ea94510-94510PRD}%
  \BibitemOpen
  \bibfield  {author} {\bibinfo {author} {\bibfnamefont {B.~B.}\ \bibnamefont
  {Brandt}}, \bibinfo {author} {\bibfnamefont {G.}~\bibnamefont {Endrodi}},
  \bibinfo {author} {\bibfnamefont {E.~S.}\ \bibnamefont {Fraga}}, \bibinfo
  {author} {\bibfnamefont {M.}~\bibnamefont {Hippert}}, \bibinfo {author}
  {\bibfnamefont {J.}~\bibnamefont {Schaffner-Bielich}},\ and\ \bibinfo
  {author} {\bibfnamefont {S.}~\bibnamefont {Schmalzbauer}},\ }\href
  {https://doi.org/10.1103/PhysRevD.98.094510} {\bibfield  {journal} {\bibinfo
  {journal} {Phys. Rev. D}\ }\textbf {\bibinfo {volume} {98}},\ \bibinfo
  {pages} {094510} (\bibinfo {year} {2018})},\ \Eprint
  {https://arxiv.org/abs/1802.06685} {arXiv:1802.06685 [hep-ph]} \BibitemShut
  {NoStop}%
\bibitem [{\citenamefont {Andersen}\ and\ \citenamefont
  {Kneschke}(2018)}]{18Andersen.Kneschke-}%
  \BibitemOpen
  \bibfield  {author} {\bibinfo {author} {\bibfnamefont {J.~O.}\ \bibnamefont
  {Andersen}}\ and\ \bibinfo {author} {\bibfnamefont {P.}~\bibnamefont
  {Kneschke}},\ }\href@noop {} {\  (\bibinfo {year} {2018})},\ \Eprint
  {https://arxiv.org/abs/1807.08951} {arXiv:1807.08951 [hep-ph]} \BibitemShut
  {NoStop}%
\bibitem [{\citenamefont {Nambu}\ and\ \citenamefont
  {Jona-Lasinio}(1961{\natexlab{a}})}]{61Nambu.Jona-Lasinio345-358PR}%
  \BibitemOpen
  \bibfield  {author} {\bibinfo {author} {\bibfnamefont {Y.}~\bibnamefont
  {Nambu}}\ and\ \bibinfo {author} {\bibfnamefont {G.}~\bibnamefont
  {Jona-Lasinio}},\ }\href {https://doi.org/10.1103/PhysRev.122.345} {\bibfield
   {journal} {\bibinfo  {journal} {Phys. Rev.}\ }\textbf {\bibinfo {volume}
  {122}},\ \bibinfo {pages} {345} (\bibinfo {year}
  {1961}{\natexlab{a}})}\BibitemShut {NoStop}%
\bibitem [{\citenamefont {Nambu}\ and\ \citenamefont
  {Jona-Lasinio}(1961{\natexlab{b}})}]{61Nambu.Jona-Lasinio246-254PR}%
  \BibitemOpen
  \bibfield  {author} {\bibinfo {author} {\bibfnamefont {Y.}~\bibnamefont
  {Nambu}}\ and\ \bibinfo {author} {\bibfnamefont {G.}~\bibnamefont
  {Jona-Lasinio}},\ }\href {https://doi.org/10.1103/PhysRev.124.246} {\bibfield
   {journal} {\bibinfo  {journal} {Phys. Rev.}\ }\textbf {\bibinfo {volume}
  {124}},\ \bibinfo {pages} {246} (\bibinfo {year}
  {1961}{\natexlab{b}})}\BibitemShut {NoStop}%
\bibitem [{\citenamefont {Vogl}\ and\ \citenamefont
  {Weise}(1991)}]{91Vogl.Weise195-272PPNP}%
  \BibitemOpen
  \bibfield  {author} {\bibinfo {author} {\bibfnamefont {U.}~\bibnamefont
  {Vogl}}\ and\ \bibinfo {author} {\bibfnamefont {W.}~\bibnamefont {Weise}},\
  }\href {https://doi.org/10.1016/0146-6410(91)90005-9} {\bibfield  {journal}
  {\bibinfo  {journal} {Prog. Part. Nucl. Phys.}\ }\textbf {\bibinfo {volume}
  {27}},\ \bibinfo {pages} {195} (\bibinfo {year} {1991})}\BibitemShut
  {NoStop}%
\bibitem [{\citenamefont {Klevansky}(1992)}]{92Klevansky649-708RMP}%
  \BibitemOpen
  \bibfield  {author} {\bibinfo {author} {\bibfnamefont {S.~P.}\ \bibnamefont
  {Klevansky}},\ }\href {https://doi.org/10.1103/RevModPhys.64.649} {\bibfield
  {journal} {\bibinfo  {journal} {Rev. Mod. Phys.}\ }\textbf {\bibinfo {volume}
  {64}},\ \bibinfo {pages} {649} (\bibinfo {year} {1992})}\BibitemShut
  {NoStop}%
\bibitem [{\citenamefont {Volkov}(1993)}]{93Volkov35-58PPN}%
  \BibitemOpen
  \bibfield  {author} {\bibinfo {author} {\bibfnamefont {M.~K.}\ \bibnamefont
  {Volkov}},\ }\href@noop {} {\bibfield  {journal} {\bibinfo  {journal} {Phys.
  Part. Nucl.}\ }\textbf {\bibinfo {volume} {24}},\ \bibinfo {pages} {35}
  (\bibinfo {year} {1993})}\BibitemShut {NoStop}%
\bibitem [{\citenamefont {Hatsuda}\ and\ \citenamefont
  {Kunihiro}(1994)}]{94Hatsuda.Kunihiro221-367PR}%
  \BibitemOpen
  \bibfield  {author} {\bibinfo {author} {\bibfnamefont {T.}~\bibnamefont
  {Hatsuda}}\ and\ \bibinfo {author} {\bibfnamefont {T.}~\bibnamefont
  {Kunihiro}},\ }\href {https://doi.org/10.1016/0370-1573(94)90022-1}
  {\bibfield  {journal} {\bibinfo  {journal} {Phys. Rept.}\ }\textbf {\bibinfo
  {volume} {247}},\ \bibinfo {pages} {221} (\bibinfo {year} {1994})},\ \Eprint
  {https://arxiv.org/abs/hep-ph/9401310} {arXiv:hep-ph/9401310 [hep-ph]}
  \BibitemShut {NoStop}%
\bibitem [{\citenamefont {Zhuang}, \citenamefont {Hufner},\ and\ \citenamefont
  {Klevansky}(1994)}]{94Zhuang.Hufner.ea525-552NPA}%
  \BibitemOpen
  \bibfield  {author} {\bibinfo {author} {\bibfnamefont {P.}~\bibnamefont
  {Zhuang}}, \bibinfo {author} {\bibfnamefont {J.}~\bibnamefont {Hufner}},\
  and\ \bibinfo {author} {\bibfnamefont {S.~P.}\ \bibnamefont {Klevansky}},\
  }\href {https://doi.org/10.1016/0375-9474(94)90743-9} {\bibfield  {journal}
  {\bibinfo  {journal} {Nucl. Phys. A}\ }\textbf {\bibinfo {volume} {576}},\
  \bibinfo {pages} {525} (\bibinfo {year} {1994})}\BibitemShut {NoStop}%
\bibitem [{\citenamefont {Lepori}\ and\ \citenamefont
  {Mannarelli}(2019)}]{19Lepori.Mannarelli96011-96011PRD}%
  \BibitemOpen
  \bibfield  {author} {\bibinfo {author} {\bibfnamefont {L.}~\bibnamefont
  {Lepori}}\ and\ \bibinfo {author} {\bibfnamefont {M.}~\bibnamefont
  {Mannarelli}},\ }\href {https://doi.org/10.1103/PhysRevD.99.096011}
  {\bibfield  {journal} {\bibinfo  {journal} {Phys. Rev. D}\ }\textbf {\bibinfo
  {volume} {99}},\ \bibinfo {pages} {096011} (\bibinfo {year} {2019})},\
  \Eprint {https://arxiv.org/abs/1901.07488} {arXiv:1901.07488 [hep-ph]}
  \BibitemShut {NoStop}%
\bibitem [{\citenamefont {He}, \citenamefont {Jin},\ and\ \citenamefont
  {Zhuang}(2006)}]{He-2006tn}%
  \BibitemOpen
  \bibfield  {author} {\bibinfo {author} {\bibfnamefont {L.}~\bibnamefont
  {He}}, \bibinfo {author} {\bibfnamefont {M.}~\bibnamefont {Jin}},\ and\
  \bibinfo {author} {\bibfnamefont {P.}~\bibnamefont {Zhuang}},\ }\href
  {https://doi.org/10.1103/PhysRevD.74.036005} {\bibfield  {journal} {\bibinfo
  {journal} {Phys. Rev. D}\ }\textbf {\bibinfo {volume} {74}},\ \bibinfo
  {pages} {036005} (\bibinfo {year} {2006})},\ \Eprint
  {https://arxiv.org/abs/hep-ph/0604224} {arXiv:hep-ph/0604224 [hep-ph]}
  \BibitemShut {NoStop}%
\bibitem [{\citenamefont {Sun}, \citenamefont {He},\ and\ \citenamefont
  {Zhuang}(2007)}]{07Sun.He.ea96004-96004PRD}%
  \BibitemOpen
  \bibfield  {author} {\bibinfo {author} {\bibfnamefont {G.-f.}\ \bibnamefont
  {Sun}}, \bibinfo {author} {\bibfnamefont {L.}~\bibnamefont {He}},\ and\
  \bibinfo {author} {\bibfnamefont {P.}~\bibnamefont {Zhuang}},\ }\href
  {https://doi.org/10.1103/PhysRevD.75.096004} {\bibfield  {journal} {\bibinfo
  {journal} {Phys. Rev. D}\ }\textbf {\bibinfo {volume} {75}},\ \bibinfo
  {pages} {096004} (\bibinfo {year} {2007})},\ \Eprint
  {https://arxiv.org/abs/hep-ph/0703159} {arXiv:hep-ph/0703159 [hep-ph]}
  \BibitemShut {NoStop}%
\bibitem [{\citenamefont {He}\ and\ \citenamefont
  {Zhuang}(2007)}]{07He.Zhuang96003-96003PRD}%
  \BibitemOpen
  \bibfield  {author} {\bibinfo {author} {\bibfnamefont {L.}~\bibnamefont
  {He}}\ and\ \bibinfo {author} {\bibfnamefont {P.}~\bibnamefont {Zhuang}},\
  }\href {https://doi.org/10.1103/PhysRevD.75.096003} {\bibfield  {journal}
  {\bibinfo  {journal} {Phys. Rev. D}\ }\textbf {\bibinfo {volume} {75}},\
  \bibinfo {pages} {096003} (\bibinfo {year} {2007})},\ \Eprint
  {https://arxiv.org/abs/hep-ph/0703042} {arXiv:hep-ph/0703042 [hep-ph]}
  \BibitemShut {NoStop}%
\bibitem [{\citenamefont {Abuki}\ \emph {et~al.}(2010)\citenamefont {Abuki},
  \citenamefont {Baym}, \citenamefont {Hatsuda},\ and\ \citenamefont
  {Yamamoto}}]{Abuki-2010jq}%
  \BibitemOpen
  \bibfield  {author} {\bibinfo {author} {\bibfnamefont {H.}~\bibnamefont
  {Abuki}}, \bibinfo {author} {\bibfnamefont {G.}~\bibnamefont {Baym}},
  \bibinfo {author} {\bibfnamefont {T.}~\bibnamefont {Hatsuda}},\ and\ \bibinfo
  {author} {\bibfnamefont {N.}~\bibnamefont {Yamamoto}},\ }\href
  {https://doi.org/10.1103/PhysRevD.81.125010} {\bibfield  {journal} {\bibinfo
  {journal} {Phys. Rev. D}\ }\textbf {\bibinfo {volume} {81}},\ \bibinfo
  {pages} {125010} (\bibinfo {year} {2010})},\ \Eprint
  {https://arxiv.org/abs/1003.0408} {arXiv:1003.0408 [hep-ph]} \BibitemShut
  {NoStop}%
\bibitem [{\citenamefont {Karsch}\ and\ \citenamefont
  {Laermann}(1994)}]{94Karsch.Laermann6954-6962PRD}%
  \BibitemOpen
  \bibfield  {author} {\bibinfo {author} {\bibfnamefont {F.}~\bibnamefont
  {Karsch}}\ and\ \bibinfo {author} {\bibfnamefont {E.}~\bibnamefont
  {Laermann}},\ }\href {https://doi.org/10.1103/PhysRevD.50.6954} {\bibfield
  {journal} {\bibinfo  {journal} {Phys. Rev. D}\ }\textbf {\bibinfo {volume}
  {50}},\ \bibinfo {pages} {6954} (\bibinfo {year} {1994})},\ \Eprint
  {https://arxiv.org/abs/hep-lat/9406008} {arXiv:hep-lat/9406008 [hep-lat]}
  \BibitemShut {NoStop}%
\end{thebibliography}%

\end{document}